\documentclass[12pt,thmsb,titlepage]{article}
\usepackage{amsmath}
\usepackage{amssymb}
\usepackage{graphicx}
\usepackage{epsf}
\usepackage{epstopdf}
\usepackage[subtle]{savetrees}

\newtheorem{theorem}{Theorem}

\newtheorem{lemma}{Lemma}
\newtheorem{proposition}{Proposition}
\newtheorem{definition}{Definition}
\newtheorem{corollary}{Corollary}

\newcommand{\captionfonts}{\footnotesize}
\makeatletter  % Allow the use of @ in command names
\long\def\@makecaption#1#2{%
  \vskip\abovecaptionskip
  \sbox\@tempboxa{{\captionfonts #1: #2}}%
  \ifdim \wd\@tempboxa >\hsize
    {\captionfonts #1: #2\par}
  \else
    \hbox to\hsize{\hfil\box\@tempboxa\hfil}%
  \fi
  \vskip\belowcaptionskip}
\makeatother   % Cancel the effect of \makeatletter

\begin{document}

\centerline{\Large{GMM is Inadmissible Under Weak Identification
%On the Inadmissibility of GMM Under Weak Identification
}}

\centerline{By Isaiah Andrews\footnote{ Harvard Department of Economics, Littauer Center M18, Cambridge, MA 02138. Email: iandrews@fas.harvard.edu.  Support from the National Science Foundation under grant number 1654234 is gratefully acknowledged.}
 and Anna Mikusheva\footnote{
Department of Economics, M.I.T., 50 Memorial Drive, E52-526, Cambridge, MA, 02142. Email: amikushe@mit.edu. We thank Jiafeng Chen, Bas Sanders, and Andrew Wang for research assistance.}} 
      
\centerline{Abstract}
 {\noindent\small{
 We consider estimation in moment condition models and show that under any bound on identification strength, asymptotically admissible (i.e. undominated) estimators in a wide class of estimation problems must be uniformly continuous in the sample moment function.  GMM estimators are in general discontinuous in the sample moments, and are thus inadmissible.  We show, by contrast, that bagged, or bootstrap aggregated, GMM estimators as well as quasi-Bayes posterior means have superior continuity properties, while results in the literature imply that they are equivalent to GMM when identification is strong.  In simulations calibrated to published instrumental variables specifications, we find that these alternatives often outperform GMM.\\
 %Hence, bagged GMM and quasi-Bayes present attractive alternatives to GMM.\\
}
Keywords: Limit Experiment, Weak Identification, Nonlinear GMM\\
JEL Codes: C11, C12, C20 }

\section{Introduction}

Generalized method of moments (GMM) estimators are ubiquitous in empirical economics, and many popular estimation methods including linear and nonlinear instrumental variables, moment-matching, and many examples of maximum likelihood can be cast as special cases.  Appropriately constructed GMM estimators are known to be efficient in large samples, in the sense of minimizing mean squared error over a large class of estimators,  provided model parameters are strongly identified (i.e. the data are sufficiently informative) and other regularity conditions hold (see Hansen 1982, Chamberlain 1987).  

Unfortunately, in many contexts of economic interest the data provide only limited information about model parameters (Mavroeidis et al. 2014, Armstrong 2016, Andrews et al. 2019).  In such cases asymptotic results assuming strong identification can be unreliable, and weak-identification approximations, which model the informativeness of the data as limited even in large samples, often provide a better description of finite-sample behavior (Staiger and Stock 1997, D. Andrews and Cheng 2012, Andrews and Mikusheva 2022).  Standard arguments for the efficiency of GMM no longer apply under weak identification, raising the question of whether GMM estimators should be used in such settings and, if not, what alternatives we should prefer.

We study asymptotic optimality under weak identification using a limit experiment derived in Andrews and Mikusheva (2022).
This limit experiment implies that there generally exists no single best estimator under weak identification, since optimizing performance over different parts of the parameter space leads to different estimators.  A minimal requirement for reasonable estimators in this setting is that they be admissible, which means there exists no alternative estimator which performs at least as well for all parameter values and strictly better for some.  Our main result shows that GMM estimators are inadmissible in decision problems where (i) the loss function is bounded and uniformly convex and (ii) identification strength is bounded.  Since these results hold for a wide class of loss functions and any bound on identification strength, they establish a strong sense in which GMM performs poorly under weak identification.

Our proof for inadmissibility is non-constructive, in the sense that it does not deliver a dominating estimator, but nonetheless suggests directions for improvement.
Specifically, we show that admissible estimators in the limit experiment must be uniformly continuous in the sample moments. To prove this result we first note that by a complete class theorem any admissible estimator under a convex loss function must be equal to the limit of Bayes decision rules for some sequence of priors.  We prove, however, that under bounds on identification strength Bayes decision rules for bounded, uniformly convex loss functions are uniformly continuous in the sample moments, so small changes in the moments lead only to  small changes in the estimator.  Since uniform continuity is preserved under limits, admissible estimators inherit this property.  GMM estimators, by contrast, change discontinuously in the sample moments when the minimizer of the sample GMM objective function is non-unique, and so fail to satisfy this necessary condition for admissibility.\footnote{The results of Guggenberger and Smith (2005) imply that under regularity conditions all Generalized Empirical Likelihood (GEL) estimators are first-order asymptotically equivalent to continuously updated GMM under weak identification.  Consequently, all of our results for GMM also apply to GEL estimators, so the latter are likewise inadmissible under weak identification.  Motivated by this equivalence, we do not separately discuss GEL approaches for the remainder of the paper.}

Motivated by the necessity of  continuity for admissibility, we next explore more continuous alternatives to GMM.
To derive positive results we focus on squared error loss, for which we show that admissible estimators much be Lipschitz in the sample moments. We discuss two estimators: first a bagged (or bootstrap aggregated) GMM estimator, and second a quasi-Bayes posterior mean. While we do not claim these estimators are admissible, we show that they have better continuity properties than GMM under weak identification, while existing results imply that both are asymptotically equivalent to GMM under strong identification and standard regularity conditions.

The first alternative estimator we discuss, bagged GMM, corresponds to the average of the GMM estimator across bootstrap realizations.  Bagging smooths the discontinuities in the GMM estimator, and we show that bagged GMM is Lipschitz in many cases. Bagged GMM has a Bayesian interpretation, corresponding to the posterior mean of the GMM estimand under an uninformative prior that does not impose correct specification of the GMM model.
%While bagging has been studied in many other settings (see e.g. Bühlmann and Yu, 2002), we are not aware of prior work on bagged GMM.  
Standard results on bootstrap bias correction (see e.g. Horowitz, 2001, Chen and Hall 2003) imply that bagged GMM is asymptotically equivalent to GMM in the strongly-identified case.\footnote{As Chen and Hall (2003) showed for estimating equation models, however, bagging is essentially the opposite of standard bias-correction, and so will increase higher-order bias in the well-identified case.}

Quasi-Bayes puts a prior on the structural parameters, treats the GMM objective as a negative log-likelihood, and combines the two to compute a quasi-posterior distribution.  This approach was initially proposed by Chernozhukov and Hong (2003) for settings where minimization is computationally intractable, and they showed that quasi-Bayes is asymptotically equivalent to GMM under strong identification.  More recently, Andrews and Mikusheva (2022) showed that quasi-Bayes arises as the limit of a sequence of Bayes decision rules under weak identification.  In the present paper, we show that quasi-Bayes posterior means are Lipschitz in the GMM objective function.  While quasi-Bayes is not in general Lipschitz in the moments, we show that it is Lipschitz in the special case where (i) the structural parameter takes only a finite number of possible values and (ii) the $J$-statistic for testing over-identifying restrictions is bounded.

We compare these estimators in simulation designs (from Andrews et al., 2019) calibrated to linear instrumental variables specifications published in the American Economic Review.  
These published specifications show substantial evidence of weak instruments, with 56 of the 121 specifications we consider having an expected first-stage F statistic smaller than 10.
We find that bagged GMM estimators for bounded target parameters typically have smaller mean squared error than their conventional counterparts, consistent with poor performance for GMM under weak identification.
We further find that the performance of quasi-Bayes depends strongly on the prior.  Specifically, quasi-Bayes estimators with a flat prior perform the worst of all estimators considered, while quasi-Bayes estimators with a novel invariant prior (motivated by invariance in the spirit of Jeffreys, 1946) are much more competitive.

Section \ref{sec- setting} describes the estimation problem we consider, the limit experiment (based on Andrews and Mikusheva 2022) in which we conduct our analysis, and defines a theoretical measure of identification strength. Section \ref{sec- admissibility} shows that admissible estimators under bounds on identification strength must be uniformly continuous in the moments, and shows that GMM fails to satisfy this condition. Section \ref{sec- alternatives} turns to alternative estimators, discussing bagged GMM in Section \ref{subsec- bagging} and quasi-Bayes in Section \ref{subsec - quasi Bayes}.  Section \ref{sec: simulations} compares the performance of these estimators in simulation.

\section{Setting}\label{sec- setting}

Consider a researcher who observes a sample of independent and identically distributed observations $X^n=\{X_i, i=1,...,n\}$ with $X_i\in \mathcal{X}.$ The distribution of the data is related to a structural parameter $\theta^*\in\Theta,$ which we assume satisfies a moment condition 
$
\mathbb{E}\left[\phi(X_i,\theta^*)\right]=0
$
for $\phi(\cdot,\cdot)$ a known $\mathbb{R}^k$-valued function of the data and parameters.

This researcher wants to choose an action $a$ from a compact and convex set of possible actions $\mathcal{A}\subseteq \mathbb{R}^p,$ where the loss from taking action $a$ when the true parameter value is $\theta^*$ is given by $L(a , \theta^*).$  We assume that $L$ is bounded, as well as continuous and uniformly convex in $a$.

\begin{definition}
Loss function $\left\{ L\left(a,\theta\right):\theta\in\Theta\right\} $
is uniformly convex in $a$ if there exists a  strictly increasing  non-negative function $\kappa:\mathbb{R_{+}\to\mathbb{R}_{+}}$ such that for all $a,a'\in\mathcal{A},$ $\theta\in\Theta$, and $t\in\left[0,1\right]$ the following inequality holds:
\[
L\left(t\cdot a+\left(1-t\right)\cdot a',\theta\right)\le
t\cdot L\left(a,\theta\right)+\left(1-t\right)\cdot L\left(a',\theta\right)-t\left(1-t\right)\kappa\left(\left\Vert a-a'\right\Vert \right).
\]
\end{definition}
A canonical example of a loss function is squared error loss for a transformation $r$ of $\theta^*$:
\begin{align}\label{eq: loss function}
L\left(a,\theta^*\right)=\left(r\left(\theta^*\right)-a\right)'\left(r\left(\theta^*\right)-a\right).
\end{align}
The loss function (\ref{eq: loss function}) satisfies our assumptions so long as the function $r(\cdot):\Theta\to\mathbb{R}^p$ is bounded. More generally, any loss of the form $L(a,\theta^*)=\ell(r\left(\theta^*\right)-a)$ with $r(\cdot)$ bounded and $\ell(\cdot)$ strictly convex satisfies our conditions, though the class of uniformly convex loss functions, and hence the class of decision problems covered by our results, is much larger.
The researcher's goal is to select a decision rule $\delta_n:\mathcal{X}^n\to\mathcal{A}$ that yields low risk, or expected loss, $\mathbb{E}[L(\delta_n(X^n),\theta^*)]$, where $\theta^*$ and the distribution of $X$ are both unknown.  Since convex loss functions are commonly used in point-estimation problems we interpret $\delta_n$ as an estimator.

GMM estimators are popular in this setting.
GMM estimates $\theta^*$ by minimizing the norm of the scaled sample moments $g_n(\theta)=\frac{1}{\sqrt{n}}\sum_{i=1}^n\phi(X_i,\theta)$, 
\[
\hat\theta_n={\arg\min}_{\theta\in\Theta}g_n(\theta)'\Xi_n(\theta)g_n(\theta),\]
for a potentially data- and parameter-dependent weighting matrix $\Xi_n(\theta)$.  
We will interpret the GMM decision rule as selecting an action $a$ through the plug-in method, $\delta_n^{GMM}(X^n)\in\arg\min_{a\in\mathcal{A}} L(a,\hat\theta_n)$.\footnote{Manski (2021) termed decision rules of this type ``as-if optimization.''}  
For instance, under squared-error loss (\ref{eq: loss function}), GMM uses the plug-in estimator $\delta_n^{GMM}(X^n)=r(\hat\theta_n)$.
Well-known asymptotic arguments (see Hansen 1982) provide conditions under which  $r(\hat\theta_n)$ is consistent for $r(\theta^*)$ and asymptotically normal as $n\to\infty.$ 
These results further establish that if $\Xi_n(\theta^*)$ is proportional to the inverse of the variance of $g_n(\theta^*)$ then the GMM estimator is asymptotically efficient under squared error loss, in the sense that $\delta_n^{GMM}$ minimizes the asymptotic risk $\lim_{n\to\infty}n\cdot \mathbb{E}[L(\delta_n(X^n),\theta^*)]$ over a large class of estimators.\footnote{Uniform integrability conditions are needed to ensure that $\lim_{n\to\infty}n\cdot \mathbb{E}[L(\delta_n(X^n),\theta^*)]$ is well-behaved.  Absent such conditions, analogous results hold for trimmed losses.}

Standard asymptotic results for GMM require, among other assumptions, that $\theta^*$ be point identified and strongly identified.  Specifically, the moment condition $\mathbb{E}\left[\phi(X_i,\theta)\right] = 0$ should be uniquely solved at $\theta^*$, and the sample moment function $g_n(\theta)$ should be well-separated from zero, asymptotically, outside infinitesimal neighborhoods of $\theta^*$.  
These point- and strong-identification assumptions are a poor fit for many economic applications, so Andrews and Mikusheva (2022) derived an alternative asymptotic efficiency theory for moment condition models with weak and partial identification.  

The weak identification asymptotics of Andrews and Mikusheva (2022) models the sample of size $n$ as generated by a distribution $P_n,$ where $P_n\to P_0$ at a $\sqrt{n}$ rate as $n\to\infty,$ and identification fails completely at $P_0$ in the sense that $\left\{\theta:E_{P_0}[\phi(X_i,\theta)]=0\right\}=\Theta.$  Hence, identification fails in the limit.\footnote{The results of Andrews and Mikusheva (2022) allow that the parameter space in the finite-sample model may be a larger set $\widetilde{\Theta}\supset\Theta.$ Any $\theta\in\widetilde\Theta\setminus\Theta$ is eventually ruled out by the data with probability one, however, so here we take $\widetilde{\Theta}=\Theta$ to simplify the exposition.}  For finite $n,$ by contrast, the model of Andrews and Mikusheva (2022) implies that $E_{P_n}[\phi(X_i,\theta)]=\frac{1}{\sqrt{n}}m(\theta)+o\left(\frac{1}{\sqrt{n}}\right),$ where $m$ satisfies $m(\theta^*)=0.$  Consequently, under mild regularity conditions the scaled sample moments converge in distribution to a Gaussian process, $g_n(\cdot)\Rightarrow g(\cdot)$ where
\begin{equation}\label{eq: GP model}
g(\cdot)\sim \mathcal{GP}(m,\Sigma),
\end{equation} 
for an unknown mean function $m$ satisfying $m(\theta^*)=0$ and a consistently estimable covariance function $\Sigma$, with $\Sigma(\theta,\tilde\theta)=\operatorname{Cov}(\phi(X_i,\theta),\phi(X_i,\tilde\theta))$. 
We treat the structural parameter $\theta^*$ as a well-defined economic quantity that may or may not be point-identified by the moment conditions.  Hence, it is meaningful to discuss the ``true'' value of $\theta^*$ even when $m$ has more than one zero so $\theta^*$ is set-identified.

For estimators, including GMM, that can be written as a function of the sample moments and the estimated covariance function $\widehat\Sigma$, it follows that $\delta_n(X^n)=\delta(g_n,\widehat\Sigma)\Rightarrow \delta(g,\Sigma)$ under mild continuity conditions.  Hence, in order to understand the large-sample behavior of these estimators it suffices to study their behavior in the limit problem.  This implies that, consistent with the results of e.g. Stock and Wright (2000), GMM estimators are inconsistent when identification is weak, with non-normal limit distributions.  Consequently, standard efficiency arguments for GMM no longer apply. 

To provide an alternative efficiency criterion, Andrews and Mikusheva (2022) further showed that (\ref{eq: GP model}) is the ``limit experiment'' in this setting. 
In the limit experiment, as in the finite-sample problem, the goal is to choose an estimator $\delta$ which now maps realizations of $g(\cdot)$ to estimates $\delta(g,\Sigma)\in \mathcal{A},$ in a way which yields a low risk  $\mathbb{E}_{m}[L(\delta(g,\Sigma),\theta^*)],$ where $\mathbb{E}_m[\cdot]$ denotes the expectation taken under \eqref{eq: GP model}. 
Andrews and Mikusheva (2022) showed that the risk in the limit experiment lower-bounds the (appropriately scaled) asymptotic risk in the original problem.
Thus, if we can find an optimal estimator  $\delta(g,\Sigma)$ in the limit experiment, the plug-in estimator $\delta_n(X^n)=\delta\left(g_n,\widehat\Sigma\right)$ will be asymptotically optimal under weak identification.

Motivated by the results of Andrews and Mikusheva (2022), the following sections focus on  properties for the limit experiment \eqref{eq: GP model}. First, however, we introduce two special cases and characterize the parameter space for the limit experiment.

\paragraph{Special Case: Linear IV}

For our first special case we consider the linear IV model.  Suppose $X_i=(Y_i,D_i,Z_i')$ for $Y_i\in\mathbb{R}$ an outcome of interest, $D_i\in\mathbb{R}$ an endogenous regressor, and $Z_i\in\mathbb{R}^k$ a vector of instruments.  The familiar linear IV estimators correspond to GMM with moment condition $\phi(X_i,\theta)=(Y_i-D_i\theta)Z_i$ and different choices of weighting matrix, for instance $\Xi_n=(\frac{1}{n}\sum Z_iZ_i')^{-1}$ for two-stage least squares.

Weak-identification asymptotics in this case correspond to weak IV asymptotics as in Staiger and Stock (1997), and model the first-stage parameter as shrinking with the sample size to ensure that  it cannot be distinguished from zero with certainty, with $\mathbb{E}_{P_n}[D_iZ_i]=\frac{1}{\sqrt{n}}\pi^*$ for a fixed vector $\pi^*$.
The $\mathbb{R}^k$-valued Gaussian process $g(\cdot)$ is linear in $\theta$, and so is fully characterized by its intercept $g(0)=\xi_{0}$ and slope $\frac{\partial}{\partial \theta}g(\theta)=-\xi_{1}$, where
\begin{equation}\label{eq: IV Model}
(\xi_0',\xi_1')'\sim N(({\pi^*}' \theta^*,{\pi^*}')',\Omega_\xi),~~\Omega_\xi=\operatorname{Var}((Z_i'Y_i,Z_i'D_i)').
\end{equation}
Intuitively, $\xi_0$ corresponds (up to a linear tranformation) to the reduced-form coefficient from regressing $Y_i$ on $Z_i$, while $\xi_1$ corresponds to the first-stage regression of $D_i$ on $Z_i$.
The two-stage least squares estimator for $\theta$ is
\begin{equation}\label{eq: TSLS}
\hat\theta={\arg\min}_{\theta\in\Theta}(\xi_0-\xi_1\theta)'\Xi(\xi_0-\xi_1\theta)=\frac{\xi_1'\Xi\xi_0}{\xi_1'\Xi \xi_1},
\end{equation}
for $\Xi=\mathbb{E}[Z_iZ_i']^{-1},$ and the corresponding GMM estimator for $r(\theta)$ under squared error loss (\ref{eq: loss function}) is $\delta(g,\Sigma)=r(\hat\theta).$ $\Box$

\paragraph{Special Case: Finite $\Theta$}

For our second special case we consider a potentially nonlinear moment condition $\phi(X_i,\theta)$ but restrict the structural parameter space to contain a finite, but potentially very large, number of points $\Theta=\{\theta_1,...,\theta_{s}\}$.  While theoretical models in economics are typically written using continuous parameterizations, computational implementation is limited by machine precision, so the case with a finite parameter space $\Theta$ is arguably a better description of empirical practice.

Weak-identification asymptotics in this setting correspond to the weak-GMM asymptotics of Stock and Wright (2000), and imply that the mean of the moments is of the same order as sampling uncertainty, $\mathbb{E}_{P_n}[\phi(X_i,\theta)]=\frac{1}{\sqrt{n}}m(\theta)+o\left(\frac{1}{\sqrt{n}}\right)$, so $\mathbb{E}_{P_n}[g_n(\theta)]=m(\theta)+o\left(1\right)$.  The limit experiment thus corresponds to observing the $sk$-dimensional normal vector $g=(g(\theta_1)',...,g(\theta_s)')'\sim N(m,\Sigma)$ for $m\in\mathbb{R}^{sk}$ and $\Sigma$ an $(sk)\times (sk)$ matrix. We assume for this example that $\Sigma$ has full rank.  The GMM estimator $\hat\theta$ for $\theta$ solves 
\[
g(\hat\theta)'\Xi(\hat\theta)g(\hat\theta)=\min\{g(\theta_1)'\Xi(\theta_1)g(\theta_1),...,g(\theta_s)'\Xi(\theta_s)g(\theta_s)\},
\]
and the GMM estimator for $r(\theta)$  under squared error loss (\ref{eq: loss function}) is $\delta^{GMM}(g,\Sigma)=r(\hat\theta).$ $\Box$

\subsection{Parameter Space for the Limit Experiment}

To complete our description of the limit experiment \eqref{eq: GP model} we need to specify the parameter space.  As in the finite sample problem, we take the parameter space for the structural parameter $\theta^*$ to be $\Theta$.   We assume that $\Theta$ is compact under a topology that renders $\Sigma(\theta,\tilde\theta)$ is continuous on the product space $\Theta\times \Theta$, that $\Sigma(\theta,\theta)$ has full rank for all $\theta$, and that $g(\cdot)$ is continuous almost surely.\footnote{ Almost sure continuity is a mild regularity condition in our setting.  As noted in Section 2.1.2 of Van der Vaart and Wellner (1996), almost-sure continuity with respect to a particular $\Sigma$-induced seminorm, and total boundedness of $\Theta$ in the same seminorm, is necessary for the sample moments to satisfy a uniform central limit theorem, $g_n(\cdot)\Rightarrow g(\cdot) \sim \mathcal{GP}(m,\Sigma)$, while Lemma 1.3.1 in Adler and Taylor (2007) implies that continuity under this seminorm is equivalent to continuity in $\theta$.}
 Andrews and Mikusheva (2022) shows that the parameter space for the functional parameter $m$ in the limit experiment is related to the reproducing kernel Hilbert space (RKHS) associated with $\Sigma,$ which we denote by $\mathcal{H}$.\footnote{For finite sets of vectors $\left\{ a_{i}\right\} _{i=1}^{s}\subset\mathbb{R}^{k}$
and $\left\{ \theta_{i}\right\} _{i=1}^{s}\subset\Theta$, consider functions of the form
$
\sum_{i=1}^{s}\Sigma\left(\cdot,\theta_{i}\right)a_{i},
$
with scalar product
$
\left\langle \sum_{i=1}^{s}\Sigma\left(\cdot,\theta_{i}\right)a_{i},\sum_{j=1}^{s^{*}}\Sigma\left(\cdot,\theta_{j}^*\right)b_{j}\right\rangle _{\mathcal{H}}=\sum_{i=1}^{s}\sum_{j=1}^{s^{*}}a_{i}'\Sigma\left(\theta_{i},\theta_{j}^{*}\right)b_{j}.
$
The RKHS $\mathcal{H}$ is the completion
of $\left\{\sum_{i=1}^{s}\Sigma\left(\cdot,\theta_{i}\right)a_{i}: a_{i}\in\mathbb{R}^{k},
 \theta_{i}\in\Theta,s<\infty\right\}$
under $\left\langle \cdot,\cdot\right\rangle _{\mathcal{H}}$.}
This is the largest set of mean functions such that we cannot tell with certainty whether a given draw $g$ was generated by $\mathcal{GP}(m,\Sigma)$ or $\mathcal{GP}(0,\Sigma)$ (see e.g. Theorem 54 of Berlinet and Thomas-Agnan, 2004).
Since $m=0$ corresponds to the case of complete non-identification of $\theta^*$,
$\mathcal{H}$ is thus the largest parameter space for $m$ such that the data never rule out complete identification failure. 

For the purposes of the present paper, it is helpful to work with a particular representation of $\mathcal{H}$.
Consider the mean-zero Gaussian process $G\sim \mathcal{GP}(0,\Sigma)$ corresponding to the noise in the moment process, $G=g-m.$  Denote by $\mathcal{C}(\Theta,\mathbb{R}^k)$ the space of $\mathbb{R}^k$-valued continuous functions on $\Theta$ with norm $\|f\|_\infty=\max_{j=1,...,k}\sup_{\theta\in\Theta}|f_j(\theta)|$ for $f\in \mathcal{C}(\Theta,\mathbb{R}^k)$, and note that under our assumptions $G\in\mathcal{C}(\Theta,\mathbb{R}^k)$ with probability one.  Let $\eta_j$ denote the point-evaluation functional at point $\theta_j,$ $\eta_j(f)=f(\theta_j),$ and note that $\mathbb{E} [G(\cdot)\eta_j(G)']=\Sigma(\cdot,\theta_j).$  If we consider linear combinations of such functionals $\eta(\cdot)=\sum_j\eta_j(\cdot)'a_j$, one can show that 
$\mathbb{E} [G(\cdot)\eta(G)]=\sum_j \Sigma(\cdot,\theta_j)a_j$ is a function in $\mathcal{H}$ and, moreover, that all functions in $\mathcal{H}$ may be written as $\mathbb{E} [G(\cdot)\eta(G)]$ for some $\eta.$

%Observe that for $\eta_j:\mathcal{C}(\Theta,\mathbb{R}^k)\to\mathbb{R}$ the linear functional such that %$\eta_j(G)=G(\theta_j)'a_j,$ $\mathbb{E} [G(\cdot)\eta_j(G)]=\Sigma(\cdot,\theta_j)a_j.$  Consequently, for $\eta(\cdot)=\sum_j\eta_j(\cdot),$ $\tilde{m}(\cdot)= \mathbb{E}[G(\cdot)\eta(G)],$ and we can represent functions in $\mathcal{H}$ in terms of the corresponding $\eta$. 

Formally, let $\mathbb{H}$ be the space of continuous linear functionals on $\mathcal{C}(\Theta,\mathbb{R}^k)$ with the norm $\|\eta\|^*=\sup_{f\in \mathcal{C}(\Theta,\mathbb{R}^k),\|f\|_\infty\leq 1}|\eta(f)|$. For each $\eta\in \mathbb{H}$ we define the Pettis integral of $\eta$ as $m_\eta(\cdot)\equiv \mathbb{E}[G(\cdot)\eta(G)]$. The RKHS $\mathcal{H}$ is the image of $\mathbb{H}$ under the 
 Pettis integral.

\begin{lemma} \label{lem: Pettis Representation} The image of $\mathbb{H}$ under the Pettis integral transformation coincides with the RKHS:
 $\mathcal{H}=\{m_\eta:\eta\in\mathbb{H}\}$.  Furthermore, the transformation is continuous with $\|m_\eta\|_\infty \le \sigma^2(G)\|\eta\|^*$, where $\sigma^2(G)=\sup_{\|\eta\|^*\le 1}\mathbb{E}[\eta(G)^2]$ is finite.
\end{lemma}
Since the mean function must be zero at the true parameter value, $m(\theta^*)=0$, we can write the parameter space in the limit experiment as
\[
\Gamma=\left\{(\theta^*,m_\eta):\theta^*\in\Theta,\eta\in\mathbb{H},m_\eta(\theta^*)=0\right\}.
\]

\paragraph{Bounding Identification Strength}

Our main result concerns parameter spaces that bound the norm of $\eta$, which we interpret as a measure of identification strength.  To understand this interpretation, consider a restricted parameter space with $\|\eta\|^*$ bounded by a positive constant $W$,
\[
\Gamma_W=\left\{(\theta^*,m_\eta):\theta^*\in\Theta,\eta\in\mathbb{H},\|\eta\|^*\le W,m_\eta(\theta^*)=0\right\}.
\]

At one extreme, if $W=0$, $\Gamma_0=\Theta\times \{0\}$ implies that $m(\theta)=0$ for all $\theta$, so $\theta^*$ is completely unidentified.  At the other extreme $\Gamma_\infty=\bigcup_W\Gamma_W=\Gamma,$ so for unrestricted $W$ we recover the original parameter space $\Gamma$.  Between these two extremes, Lemma \ref{lem: Pettis Representation} shows that for any $(\theta^*,m)\in\Gamma_W,$ $\|m\|_\infty\le\sigma^2(G)W.$  Since we observe only a noisy measure of $m$, $g(\cdot)\sim\mathcal{GP}(m,\Sigma),$ bounds on $\|m\|_\infty$ limit the ease with which we can distinguish $m(\theta)$ from $0$ for any $\theta$ value and so limit how informative the data can be about $\theta^*$.  Thus, we can interpret $\Gamma_W$ as a parameter space which imposes a uniform upper bound on the strength of identification.

In infinite-dimensional settings bounds on $\|\eta\|^*$ imply upper, but not in general lower, bounds on $\|m\|_\infty.$
Consequently, bounding $\|\eta\|^*$ is more restrictive than bounding $\|m\|_\infty,$ so 
weak identification neighborhoods defined using $\|\eta\|^*$ are ``weaker'' in that sense. The choice of the norm $\|\cdot\|^*$ as our measure of identification strength is natural, since we show in the proof of our main result that this
is the norm that translates changes in the moments $\|g-g'\|_\infty$ to changes in the likelihood. In the finite dimensional case, however, the choice of norm is unimportant since the norms $\|\cdot\|^*$ and $\|\cdot\|_\infty$ are equivalent.

\paragraph{Finite-Dimensional Limit Experiments}
In many cases of empirical interest the limit experiment is finite-dimensional, in the sense that $g(\cdot)$ can be written as a function of a finite-dimensional normal random vector or, equivalently, that the covariance function $\Sigma$ has a finite number of nonzero eigenvalues.
\begin{definition}
The limit experiment is finite-dimensional if the covariance function $\Sigma$ has finitely many nonzero eigenvalues.
\end{definition}
Most of our results apply to both finite- and infinite-dimensional limit experiments, but the interpretation of some conditions is simpler in the finite-dimensional case.
Finite-dimensional limit experiments can arise in many ways, for instance because the support $\mathcal{X}$ of the data is finite, because the moments are additively or multiplicatively separable in the data, $\phi(X_i,\theta)=\phi_{1}(X)\phi_{2}(\theta)$ for $\phi_1(\theta)\in\mathbb{R}^{k\times d}$ and $\phi_2(x)\in\mathbb{R}^{d}$, or because the parameter space is finite.  Whatever the source of finite dimension, our bounds on identification strength are particularly easy to interpret in this case. Specifically, since all norms are equivalent on finite-dimensional spaces, there exists a ($\Sigma$-dependent) constant $\lambda$ such that $\lambda ^{-1}\|m_\eta\|_\infty\le \|\eta\|^* \le \lambda \|m_\eta\|_\infty,$ so bounds on $\|\eta\|^*$ not only imply, but are also implied by, bounds on $\|m\|_\infty$.  
 
\paragraph{Special Case: Linear IV (continued)}

Recall that in the linear IV model $g(\theta)=\xi_0-\xi_1\theta$ for $(\xi_0,\xi_1)$ a Gaussian vector in $\mathbb{R}^{2k}$.  This is therefore a finite-dimensional setting.  The mean function is $m(\theta)=\pi^*(\theta^*-\theta),$ so bounding $\|m\|_\infty$ is equivalent to bounding the first-stage $\pi^*$.  Consequently, for $\pi^*_\eta$ the first-stage  implied by $\eta$ and $\|\pi_\eta^*\|$ its Euclidean norm, there exists a constant $\lambda^*$ such that $\lambda ^{*-1}\|\pi^*_\eta\|\le\|\eta\|^* \le\lambda^*\|\pi_\eta^*\|,$ and bounding $\|\eta\|^*$ is equivalent to bounding the first-stage coefficient $\pi^*$. $\Box$

\paragraph{Special Case: Finite $\Theta$ (continued)}

In this example the process $g$ reduces to a Gaussian vector in $\mathbb{R}^{sk},$ so this is again a finite-dimensional case.  Thus, there exists a constant $\lambda^*$ for which $\lambda^{*-1}\|m_\eta\|_\infty\le\|\eta\|^*\le \lambda^{*}\|m_\eta\|_\infty,$ and bounding identification strength in terms of $\|\eta\|^*$ is equivalent to bounding $\|m\|_\infty$, the maximal deviation of the moments from zero.  $\Box$

\section{Admissibility}\label{sec- admissibility}

Recall that the limit experiment corresponds to observing $g\sim\mathcal{GP}(m,\Sigma),$ where $(\theta^*,m)\in \Gamma.$
  The researcher aims to choose an estimator $\delta$ that yields a low risk $\mathbb{E}_{m}[L(\delta(g,\Sigma),\theta^*)]$ for the loss function $L$,  
  where since $\Sigma$ is known in the limit experiment we abbreviate $\delta(g,\Sigma)=\delta(g)$ going forward.
 Unfortunately there is not a uniformly best estimator in this setting, as minimizing risk at different parameter values $(\theta^*,m),({\theta^*}',m')\in \Gamma$ usually leads to distinct estimators $\delta$ and $\delta'$. It is however without loss of performance to limit attention to the set of admissible estimators.
  
 \begin{definition}
  An estimator $\delta$ is dominated on $\widetilde\Gamma\subseteq\Gamma$ if there exists another estimator $\delta'$ such that $\mathbb{E}_{m}[L(\delta'(g),\theta^*)]\le \mathbb{E}_{m}[L(\delta(g),\theta^*)]$ for all $(\theta^*,m)\in \widetilde\Gamma$, with a strict inequality for some $(\theta^*,m)\in \widetilde\Gamma$.  The estimator $\delta$ is admissible on $\widetilde\Gamma$ if it is not dominated on $\widetilde\Gamma$.  
 \end{definition} 

An estimator is admissible if its performance, measured in terms of risk,  cannot be uniformly improved.  Since no admissible estimator  dominates any other, selecting from among sets of admissible estimators requires taking a stand on how we value performance over different regions of the parameter space, for instance by specifying a prior and considering Bayes estimators as in Andrews and Mikusheva (2022).  In the present paper we set a more modest goal, and aim to provide necessary conditions for admissibility under bounds on identification strength. Our main technical contribution is to establish a close connection between the set of admissible estimators under bounded identification strength and the set of estimators that are uniformly continuous in $g$.

% \begin{definition}
% Loss function $\left\{ L\left(a,\theta\right):\theta\in\Theta\right\} $
% is uniformly convex in the first argument if there exists a  strictly increasing  non-negative function $\kappa:\mathbb{R_{+}\to\mathbb{R}_{+}}$ such that for all $d,d'\in\mathcal{D},\theta\in\Theta,t\in\left[0,1\right]$ the following inequality holds:
% \[
% L\left(t\cdot d+\left(1-t\right)\cdot d',\theta\right)\le
% t\cdot L\left(a,\theta\right)+\left(1-t\right)\cdot L\left(d',\theta\right)-t\left(1-t\right)\kappa\left(\left\Vert d-d'\right\Vert \right).
% \]
% \end{definition}

\begin{definition}
An estimator $\delta$ is almost-surely uniformly continuous with modulus of continuity $\tilde\kappa(\cdot)$, where $\tilde\kappa(\cdot)$ is continuous at zero and $\tilde\kappa(0)=0$,  if there exist another estimator $\delta^*$ such that $\delta(g)=\delta^*(g)$ for almost every $g$ and  $\|\delta^*(g)-\delta^*(g')\|\le \tilde\kappa\left( \|g-g'\|_\infty\right)$ for all $g,g'$ in the support of the process $\mathcal{GP}(0,\Sigma)$.  
\end{definition}

\begin{theorem}\label{thm: ae continuity} Assume that the loss function $L(a,\theta)$ is bounded, continuous and uniformly convex in its first argument. Assume that an estimator $\delta$ is admissible on $\widetilde\Gamma$, where $\widetilde\Gamma\subseteq\Gamma_W$ for $W<\infty$. Then $\delta$ is almost-surely uniformly continuous
with modulus of continuity $\tilde\kappa(x)=\kappa^{-1}\left(2\bar{L}W\cdot x\right)$ for $\bar{L}=\sup_{a,\theta}L(a,\theta).$
\end{theorem}

The proof of Theorem \ref{thm: ae continuity} builds on Theorem 2 of Andrews and Mikusheva (2022), which is itself a minor extension of a result from Brown (1986).  That result, reproduced in the appendix for completeness, shows that for convex loss functions admissible estimators must be the (almost everywhere) pointwise limit of Bayes decision rules  for finitely-supported priors.
Going beyond this result, we then prove that under bounded identification strength small changes in the moments $g$ lead to only small changes in the posterior probability of different $\theta$ values.\footnote{The choice of the norm $\|\cdot\|^*$ as our measure of identification strength is important for this step, since it allows us to bound the change in the likelihood between moment realizations $g$ and $g'$ in terms of $\|g-g'\|_\infty$.} We further prove that for a bounded loss the posterior expected risk function is Lipschitz in $g$. Uniform convexity of the loss function allows us to translate this last property into uniform continuity of Bayes decision rules. Uniform continuity is preserved by pointwise limits, so the conclusion of the theorem follows.

If we consider the special case of squared error loss (\ref{eq: loss function}), the previous result can be strengthened to show that admissible decision rules must be Lipschitz in $g$.

\begin{definition}
An estimator $\delta$ is almost-surely Lipschitz with Lipschitz constant $K$ if there exists another estimator $\delta^*$ such that $\delta(g)=\delta^*(g)$ for almost every $g$ and $\|\delta^*(g)-\delta^*(g')\|\le K \|g-g'\|_\infty$ for all $g,g'$ in the support of the process $\mathcal{GP}(0,\Sigma)$.  
\end{definition}

\begin{theorem}\label{thm: ae continuity1} Assume that an estimator $\delta$ is admissible on $\widetilde\Gamma$, where $\widetilde\Gamma\subseteq\Gamma_W$ for $W<\infty$ for squared error loss as defined in (\ref{eq: loss function}). Then $\delta$ is almost-surely Lipschitz with Lipschitz constant $K=\bar{r}\sqrt{p} W,$ where $\bar{r}=\sup_\theta\|r(\theta)\|$.
\end{theorem}

%It is important to emphasize that the set of admissible estimators depends on the set of parameter values $\widetilde{\Gamma}$ over which the performance is evaluated, and that the set of admissible estimators  is in general not monotone in $\widetilde{\Gamma}$.  That is, if we enlarge $\widetilde{\Gamma}$ the set of admissible estimators may lose some estimators but gain others.  Motivated by this fact, Theorems \ref{thm: ae continuity} and \ref{thm: ae continuity1} consider the set of estimators which are admissible for some set $\widetilde{\Gamma}$ that obeys a numerical bound $W$ on identification strength. 

While Theorems \ref{thm: ae continuity} and \ref{thm: ae continuity1} translate numerical bounds on identification strength to numerical bounds on the modulus of continuity or Lipschitz constant, selecting a value of $W$ for a given application seems unappealing.  Similarly, different researchers may prefer different loss functions.  We next provide a necessary condition for admissibility under \emph{any} bound on identification strength and \emph{any} uniformly convex loss.

\begin{corollary}\label{cor: Continuous}
If $\delta$ is not almost-surely uniformly continuous, then it is inadmissible on $\widetilde{\Gamma}$ for all $\widetilde{\Gamma}$ with bounded identification strength (that is, all $\widetilde{\Gamma}$ with $\widetilde{\Gamma}\subseteq\Gamma_W$ for some $W<\infty$) and all bounded, continuous, and uniformly convex loss functions.
\end{corollary}

Corollary \ref{cor: Continuous} states that under any bound on the strength of identification, no matter how large, and any uniformly convex loss, admissible estimators are uniformly continuous in the moment process $g$, so small changes in the realized sample moments (measured in the supremum norm $\|\cdot\|_\infty$) can induce only small changes in the estimate.  Hence, if we observe that a given estimator fails to be uniformly continuous, it is unappealing in a very wide variety of decision problems.
While uniform continuity may seem a minimal requirement, we show in the next section that GMM estimators do not have this property.

We show in Appendix \ref{sec: discontinuity example} that bounded identification strength is crucial for the uniform continuity property.  There, we provide an example with an unrestricted parameter space $\Gamma$ where the limit of Bayes posterior means is a step function, and thus discontinuous. 
%While limit-of-Bayes property is necessary for admissibility, however, it is not in general sufficient.  Hence, this example does not establish one way or the other whether all admissible estimators are Lipschitz without restriction on the identification strength. That said, if there exist estimators that are admissible under unbounded identification strength but that are not Lipschitz, justifications for such estimators necessarily involve regions of the parameter space $\Gamma$ where identification is arbitrarily strong (that is, $(\theta^*,m)$ pairs which are excluded from $\Gamma_W$ for all finite $W$).  Given that datasets encountered in practice, however large, are always finite, and so contain only a limited amount of information about $\theta^*$, it is unclear that such justifications should be practically compelling, and it seems potentially appealing to limit attention to Lipschitz estimators. 

\subsection{Inadmissibility of GMM}\label{sec: inadmissibility}

While it is known that appropriately constructed GMM estimators are efficient in large samples when the model is strongly identified, this result breaks down under weak identification.
We next show that GMM estimators are not generally uniformly continuous, and so are inadmissible under any bound on the strength of identification and for a wide variety of convex loss functions.  GMM estimators in the limit experiment take the form 
\begin{equation}\label{eq: GMM estimator}
\delta^{GMM}(g,\Sigma)=\arg\min_{a\in\mathcal{A}} L(a,\hat\theta),~~~\hat\theta\in \arg\min_{\theta\in\Theta}g(\theta)'\Xi(\theta)g(\theta),
\end{equation}
where $\Xi(\theta)$ is a deterministic weight function, corresponding to the probability limit of $\Xi_n(\theta)$.  If there are multiple points where the second minimum is achieved, we assume that $\hat\theta$ applies some selection rule.

GMM estimators in the limit experiment are invariant to the scale of $g$.\footnote{To be precise, GMM estimators are scale-invariant so long as the rule for selecting from a non-unique argmin is likewise invariant.}
\begin{definition}
An estimator $\delta$ is scale-invariant if $\delta(c\cdot g,\Sigma)=\delta(g,\Sigma)$ for all $g$ and all $c>0.$
\end{definition}
This scale invariance is important for our purposes, since scale-invariant estimators in our setting are uniformly continuous if and only if they are constant.
\begin{lemma}\label{lem: non-Lipschitz}
Let $\delta$ be a scale-invariant estimator. If $\delta$ is almost-surely uniformly continuous with the modulus of continuity $\tilde\kappa(\cdot)$, where $\tilde \kappa ( \cdot )$ is continuous at zero and $\tilde\kappa(0)=0$, then  there exists $a^*\in\mathcal{A}$ such that $\delta(g,\Sigma)=a^*$ almost surely.
\end{lemma}
GMM estimators $\delta^{GMM}$ are scale-invariant and non-constant, so Lemma \ref{lem: non-Lipschitz} implies that $\delta^{GMM}$ is not uniformly continuous. Hence, by Corollary \ref{cor: Continuous}, $\delta^{GMM}$ is inadmissible under bounded identification strength.  
The source of this inadmissibility is intuitive, namely that small changes in data can cause the GMM estimator to jump discontinuously.
Our weakly identified setting is important for this result, and GMM estimators are known to be asymptotically admissible in many strongly-identified problems.  Appendix \ref{sec: strong ID} discusses why Lemma \ref{lem: non-Lipschitz} does not apply in the strongly identified case.

\paragraph{Special Case: Linear IV (continued)}

In this example, $\delta$ is continuous in $g(\cdot)$ if and only if it is continuous in $(\xi_0,\xi_1),$ and the two-stage least squared estimator \eqref{eq: TSLS} is discontinuous when the first-stage estimate is zero, $\xi_1=0.$ This is consistent with the intuition that instrumental variables estimation is badly behaved when the instrument is irrelevant.  

Interestingly, if we use other instrumental variables estimators (with multiple instruments, $k>1$) we may encounter additional points of discontinuity.  For instance the limited information maximum likelihood estimator corresponds to GMM  with weighting matrix
$\Xi(\theta)=(\sigma_u^2-2\sigma_{uv}\theta+\sigma_{v}^2\theta^2)^{-1}\mathbb{E}[Z_iZ_i']^{-1}$
for $\sigma^2_{u},$ $\sigma^2_{v},$ and $\sigma_{uv}$  the residual variances and covariance from regressing $(Y,D)$ on $Z.$  The resulting estimator $\hat\theta$ is discontinuous at $\xi_1=0,$ but also at $(\xi_0,\xi_1)$ where (i) the OLS and two-stage least squares estimates coincide and (ii) the reduced-form $R^2$ coefficient exceeds the first-stage $R^2$, which may be interpreted as a sign of model misspecification -- see Andrews (2019). $\Box$ %\footnote{The unbiased IV estimators proposed by Andrews and Armstrong (2017) are continious, but not Lipschitz, in the just-identified ($k=1$) case.  In the over-identified case, their continuity properties depend on the weights used.} 

%\paragraph{Special Case: Finite $\Theta$ (continued)}

%In this case the GMM estimator $\hat\theta$ is a step function, and is discontinuous wherever ${\arg\min}_{\theta\in\Theta}g(\theta)'W(\theta)g(\theta)$ is non-unique.  The discrete nature of the parameter space in this setting is particularly unfavorable for continuity of the GMM estimator.

\section{Alternative Estimators}\label{sec- alternatives}

In the last section we showed that GMM estimators are inadmissible under bounds on identification strength.  Unfortunately our proof is non-constructive, and yields no characterization for a dominating estimator.  The \emph{reasons} for GMM's inadmissibility are nonetheless instructive, and suggest a route to more reasonable estimators. The source of GMM's inadmissibility is that it depends only on the minimizer of the GMM objective.  This results in the scale-invariance discussed in the last section which, in turn, implies that the GMM estimate is not continuous in the moments $g.$  

In this section we study two estimators which depend on the moments in a more continuous way, the first based on bagging or bootstrap aggregation, and the second based on quasi-Bayes.  In order to provide constructive results we focus on squared error loss (\ref{eq: loss function}), so Bayes estimators correspond to posterior means, and Theorem \ref{thm: ae continuity1} implies that admissible estimators must be Lipschitz.   Both  bagged GMM and Quasi-Bayes are continuous in the moments, and both are Lipschitz under additional conditions. 

While we show that bagged GMM and quasi-Bayes have better continuity properties than GMM when identification is weak, and existing theoretical results imply that both are efficient when identification is strong,  their admissibility under weak identification is an open question.  An alternative approach, which guarantees admissibility under weak identification, is to 
report Bayes posterior means based on full-support priors on $\Gamma_W$.  To implement this approach in practice, however, one would have to take a stand on the identification strength bound $W,$ which seems unlikely to be appealing.  Moreover, in the infinite-dimensional case it is not obvious to us how to construct such priors or compute the resulting posteriors.  Hence, while a fully Bayesian approach guarantees admissibility, we expect Quasi-Bayes and bagged GMM to be more attractive options in most applications.

%\textbf{FIXME:In this section we consider the quadratic loss function only. For this loss function Theorem \ref{thm: ae continuity1} implies that admissible estimators should be Lipschitz in the realization of the data. The Bayes decision rules corresponds to the posterior means. }

\subsection{Bagged GMM}\label{subsec- bagging}

One way to ensure continuity is to directly smooth the GMM estimator.  For instance, we can average an estimator across bootstrap draws, yielding a bagged, or bootstrap aggregated, estimator.  Bühlmann and Yu (2002) show that bagging can reduce both bias and variance when estimators are unstable, in the sense of being sensitive to small changes in the data.  The instability of GMM under weak identification suggests that its performance might also be improved by bagging. 

To formally introduce the bagged GMM estimator, again consider the limit experiment where we observe a single draw of the moment process $g\sim\mathcal{GP}(m,\Sigma)$.  For an estimator $\delta(g)$ (which need not be GMM) let us draw independent Gaussian noise $\zeta\sim\mathcal{GP}(0,\Sigma)$ and define the bagged version of $\delta$ as the average of $\delta(g+\zeta)$ over noise realizations,
$$
\delta^{B}(g)=\mathbb{E}\left[\delta(g+\zeta)|g\right].
$$
We interpret $\delta^{B}(g)$ as a bagged estimator because the distribution of $g^*=g+\zeta$ given $g$ is exactly the asymptotic distribution of the moments across bootstrap replications, conditional on the initial data delivering moments $g$ (see e.g. Section 3.6 and Van der Vaart and Wellner 1996).  Hence, $\delta^{B}(g)$ corresponds to the (asymptotic analog of the) average of $\delta(\cdot)$ across bootstrap draws.

We formalize the connection between bagging and smoothing in our setting by showing that in finite-dimensional limit experiments all bagged estimators are Lipschitz.
\begin{proposition}\label{prop: bagged estimator}
If the limit experiment is finite-dimensional, then for any estimator $\delta(\cdot)$ with range contained in $\mathcal{A}$ the bagged estimator $\delta^B(g)$ is Lipschitz.
\end{proposition}
Proposition \ref{prop: bagged estimator} implies, in particular, that for $\delta^{GMM}(g)$ the GMM estimator as defined in \eqref{eq: GMM estimator}, the bagged GMM estimator $\delta^{BGMM}(g)=\mathbb{E}[\delta^{GMM}(g+\zeta)|g]$ satisfies the global continuity property required by Corollary \ref{cor: Continuous}.  On an intuitive level this estimator ``averages out'' the discontinuities of GMM, resulting in a Lipschitz estimator.\footnote{The same is true for the whole family of estimators $\delta^{BGMM}_\tau(g)=\mathbb{E}\left[\delta(g+\tau\cdot\zeta)|g\right]$ for $\tau>0$.  However, values $\tau\neq 1$ complicate the bootstrap interpretation, as well as the Bayesian interpretation discussed below, so we focus on the case with $\tau=1$.}  A practical limitation of the bagged GMM estimator is that it requires repeatedly minimizing the GMM objective function to compute $\delta^{GMM}(g+\zeta)$.  In settings where minimization is difficult this can make computing the bagged estimator costly.  If, on the other hand, a researcher is already using the bootstrap then the incremental cost of computing the bagged estimator is essentially zero.

To illustrate how bagging smooths the GMM estimator, we return to our examples.

\paragraph{Special Case: Finite $\Theta$ (continued)}

In this special case the bagged GMM estimator for $r(\theta)$ can be written as a weighted average across the possible values of $r(\theta).$
In particular, for $\hat\theta(g+\zeta)$ the GMM estimate based on moments $g+\zeta,$  we can write the bagged GMM estimator as
$
\delta^{BGMM}(g)=\sum_{\theta\in\Theta} r(\theta) Pr\left\{\hat\theta(g+\zeta)=\theta|g\right\}.
$
Note, however, that the probability $Pr\left\{\hat\theta(g+\zeta)=\theta|g\right\}$ is simply the probability (conditional on $g$) that the collection of correlated random variables $\{(g(\theta)+\zeta(\theta))'\Xi(\theta)(g(\theta)+\zeta(\theta)):\theta\in\Theta\}$ achives its minimum at a particular $\theta$.  One can show that these probabilities are Lipschitz in $g$, from which it is immediate that $\delta^{BGMM}(g)$ is Lipschitz as well. $\Box$

\paragraph{Special Case: Linear IV (continued)}

Recall that the limit experiment for linear IV reduces to observing the jointly normal random vector $\xi=(\xi_0',\xi_1')'\sim N(({\pi^*}'\theta^*,{\pi^*}')',\Omega_\xi)$, corresponding to the intercept and (negative) slope of $g$.  We can correspondingly define
 $\nu=({\nu_0}',{\nu_1}')\sim N(0,\Omega_\xi)$ as the intercept and (negative) slope of $\zeta\sim \mathcal{GP}(0,\Sigma).$ The two-stage least squares estimate of $\theta$ for moment realization $g+\zeta$ is thus
\[\hat\theta(\xi+\nu)=\frac{(\xi_1+\nu_1)'\Xi(\xi_0+\nu_0)}{(\xi_1+\nu_1)'\Xi (\xi_1+\nu_1)},
\]
while the two-stage least squares estimate of $r(\theta)$ is $r\left(\hat\theta(\xi+\nu)\right)$.  The bagged two-stage least squares estimator is then
\[
\delta^{BTSLS}(g)=E\left[r\left(\hat\theta(\xi+\nu)\right)|g\right]=\int r\left(\hat\theta(\xi+\nu)\right)f(\nu)d\nu=\int r\left(\hat\theta(\nu)\right)f(\nu-\xi)d\nu,
\]
for $f(\cdot)$ the $N(0,\Omega_\xi)$ density.  Note, however, that $f(\cdot)$ is Lipschitz, so since $r\left(\hat\theta(\cdot)\right)$ is bounded,
$\delta^{BTSLS}(g)$ is likewise Lipschitz. $\Box$

The bagged GMM estimator also has a Bayesian interpretation.  In the finite-dimensional case the mean function $m$ is simply a finite-dimensional vector.  For a flat (improper) prior on $m$, the posterior distribution on $m$ after observing $g$ corresponds to a $\mathcal{GP}(g,\Sigma)$ distribution, which is precisely the distribution of $g+\zeta$ conditional on $g$.  Note, however, that the flat prior on $m$ allows the possibility that $m(\theta)\neq0$ for all $\theta$ and so does not impose correct specification of the GMM model.  This raises the question of how to define the object of interest when GMM is misspecified. 
One approach is to focus on the GMM estimand or pseudo-true value $\theta^*(m)=\arg\min_{\theta\in\Theta}m(\theta)'\Xi(\theta)m(\theta),$ which minimizes the population analog of the GMM objective.  The bagged GMM estimator then corresponds to the posterior mean of $r(\theta^*(m))$ under the flat prior.

\subsection{Quasi-Bayes}\label{subsec - quasi Bayes}

We could also take a more overtly Bayesian approach. For a prior $\pi$ on $\Theta,$ the quasi-Bayes posterior mean of $r(\theta)$ in the limit experiment is
\begin{equation}\label{eq: quasi-Bayes}
\delta_{\pi}^{QB}\left(g\right)=\int r\left(\theta\right)\frac{\exp\left(-\frac{1}{2}Q(\theta|g)\right)}{\int\exp\left(-\frac{1}{2}Q(\theta|g)\right)d\pi(\theta)}d\pi\left(\theta\right),
\end{equation}
where $Q(\theta|g)=g\left(\theta\right)'\Sigma\left(\theta,\theta\right)^{-1}g\left(\theta\right)$ is the continuously updated GMM objective function.  
This estimator corresponds to the posterior mean after updating $\pi(\theta)$ based on ``log-likelihood'' $-\frac{1}{2}Q(\theta|g),$ and was initially suggested by Chernozhukov and Hong (2003).  Since $Q(\theta|g)$ is not in general the likelihood of the researcher's model the interpretation of $\delta_{\pi}^{QB}\left(g\right)$ from a strict Bayesian perspective may not be obvious, but Andrews and Mikusheva (2022) showed that this estimator arises as the limit of a sequence of Bayes posterior means for proper priors. Unlike the priors underlying bagged GMM, the priors that give rise to quasi-Bayes impose correct specification of the GMM model.
See Chernozhukov and Hong (2003) and Andrews and Mikusheva (2022) for further discussion, as well as asymptotic results under both strong and weak identification.

A key feature of the quasi-Bayes approach for our purposes is that it takes a weighted average of $r(\theta)$ over the parameter space $\Theta$, weighting by $\frac{\exp\left(-\frac{1}{2}Q(\theta|g)\right)}{\int\exp\left(-\frac{1}{2}Q(\theta|g)\right)d\pi(\theta)}d\pi\left(\theta\right)$. It follows from this structure that quasi-Bayes is Lipschitz in the GMM objective function.
 %This directly incorporates the level of the GMM objective function, so the minimizer receives only slightly more weight than near-minimizers.  Hence, if the objective function is close to flat over some region of the parameter space and high outside of this region, the quasi-Bayes estimator will correspond to a weighted average over the flat region, and will not be very sensitive to the precise location of the global minimum.  If instead the GMM objective has a well-separated minimum, the quasi-Bayes estimator will be close to the argmin.  This structure implies that the quasi-Bayes estimator is continuous, and indeed Lipschitz, in the GMM objective function.
\begin{lemma}\label{lem: QB lipchitz in Q} Quasi-Bayes is Lipschitz in the GMM objective function $Q$:
\[
\left\|\delta^{QB}\left(g\right)-\delta^{QB}\left(g'\right)\right\|\le K\left\Vert Q\left(\cdot|g\right)-Q\left(\cdot|g'\right)\right\Vert _{\infty},\]
where $K=\frac{1}{2}\bar{r}\sqrt{p}.$
\end{lemma}
Unfortunately, $Q(\cdot|g)$ is continuous but not Lipschitz in the moments $g$.  Consequently, the Lipschitz continuity required by Theorem \ref{thm: ae continuity1} does not follow from Lemma \ref{lem: QB lipchitz in Q}.  Indeed, while quasi-Bayes is continuous in $g$, it is not in general Lipschitz.\footnote{Since Andrews and Mikusheva (2022) showed that quasi-Bayes emerges as the limit of a sequence of Bayes posterior means, it may be surprising that it does not satisfy a necessary condition for admissibility under bounded identification strength.  The sequences of priors underlying quasi-Bayes imply, however, that $\|m\|_\infty\to_p\infty$ and so correspond to the case of unbounded identification strength.} 

\paragraph{Special Case: Finite $\Theta$ (continued)} Suppose that the parameter space consists of just two points, $\Theta=\{0,1\},$ that we have a one-dimensional moment condition $(k=1),$ and that $\Sigma=I_2$. Consider the quasi-Bayes estimator using a prior $\pi$ that puts weight $\frac{1}{2}$ on each parameter value. The quasi-Bayes estimator of $\theta$ is
\[
\delta^{QB}_\pi(g)=\frac{\exp\left(-\frac{1}{2}g\left(1\right)^{2}\right)}{\exp\left(-\frac{1}{2}g\left(0\right)^{2}\right)+\exp\left(-\frac{1}{2}g\left(1\right)^{2}\right)}=\frac{1}{1+\exp\left(\frac{1}{2}g\left(1\right)^{2}-\frac{1}{2}g\left(0\right)^{2}\right)}.
\]
While this estimator is differentiable in   $(g(0),g(1))$, it is not Lipschitz.
Indeed,
\[
\left.\frac{\partial\delta^{QB}_\pi(g)}{\partial g(0)}\right|_{g(0)=g(1)}=\frac{g(0)}{4},
\]
which exceeds any finite constant for large values of both  $g(0)$ and $g(1)$.  Intuitively, when both $g(0)$ and $g(1)$ are large, $\delta^{QB}_\pi(g)$ behaves like $\delta(g)={\arg\min}_{\theta\in\{0,1\}} g(\theta)^2.$ $\Box$

An interesting feature of this example is that the non-Lipschitz behavior of the quasi-Bayes estimator appears for realizations of $g$ which suggest misspecification of the model.  Specifically, the GMM model with parameter space $\Theta=\{0,1\}$ requires that either $m(0)=0$ or $m(1)=0.$ Hence, under the model the distribution of ${\min}_{\theta\in\{0,1\}} g(\theta)^2$ is bounded by a $\chi^2_1,$ and data realizations with both $g(0)$ and $g(1)$ large are highly unlikely.  This suggests that if we limit attention to data realizations which appear consistent with the model the quasi-Bayes estimator may be Lipschitz.  The following result shows that this is the case provided $\Theta$ is finite and $\pi$ has full support.

\begin{proposition}\label{prop: quasi-bayes lipschitz}
Assume that the parameter space is finite, $|\Theta|<\infty$. For $C>0$ define
\[
\mathcal{G}_{C}=\left\{ g:\inf_{\theta\in\Theta}Q\left(\theta|g\right)\le C\right\}.
\] 
If $\pi$ has support $\Theta,$ then the quasi-Bayes estimator $\delta^{QB}_\pi(g)$ is Lipschitz
in $g$ on $\mathcal{G}_{C}$.
\end{proposition}

The minimized GMM objective $Q\left(\theta|g\right)$ is often  termed a $J$-statistic, and researchers commonly reject correct specification of the model when this statistic exceeds a threshold.  Under the assumption of correct specification we have
$
\lim_{C\to\infty}\inf_{\gamma\in\Gamma}\mathbb{P}_{\gamma}\left\{ g\in\mathcal{G}_{C}\right\}=1,
$
so moment realizations $g\not\in\mathcal{G}_{C}$ have low probability under all data generating processes consistent with the GMM model.  Hence, for finite $\Theta,$ quasi-Bayes is Lipschitz over data realizations such that the GMM model is not rejected.

%\textbf{FIXME: suggestion- to cut next paragraph}
%Outside of the finite $\Theta$ case, bounding the $J$-statistic is insufficient to ensure that quasi-Bayes is Lipschitz, but one may modify quasi-Bayes to make it Lipschitz.\footnote{Since Andrews and Mikusheva (2022) show that quasi-Bayes emerges as the limit of a sequence of Bayesian estimators, it may be surprising that it does not satisfy the necessary condition for admissibility under bounded identification strength.  The  diffuse priors underlying quasi-Bayes imply, however, that $\|m\|_\infty\to_p\infty$, and so correspond to the case of unbounded identification strength.}  For $q>0$, if we define the (Huber 1964-type) function $f_q(\cdot)$ by
%\[
%f_{q}\left(x\right)=\begin{cases}
%x & \text{if \ensuremath{x < q}}\\
%\sqrt{x}\sqrt{q} & \text{if \ensuremath{x \ge q}}
%\end{cases},
%\]
%then $f_q(Q(\cdot|g))$ is Lipschitz in $g$.  Hence, if we define a modified quasi-Bayes estimator $\delta^{QB}_{\pi,q}(g)$ which replaces $Q(\theta|g)$ in \eqref{eq: quasi-Bayes} by $f_q(Q(\cdot|g))$, the same argument used to prove Lemma \ref{prop: quasi-bayes lipschitz} implies that $\delta^{QB}_{\pi,q}(g)$ is Lipschitz in $g$.  The decision-theoretic motivation for $\delta^{QB}_{\pi,q}(g)$ is unclear, however, so in finite-dimensional settings where a fully Lipschitz estimator is desired we prefer the bagging approach discussed next.

\subsubsection{Default Priors}

To apply the quasi-Bayes approach (unlike for bagged GMM) we must explicitly specify a prior $\pi(\theta)$. From a subjective Bayesian perspective the prior $\pi$ on the GMM
parameter $\theta$ should reflect the researcher's beliefs
about the structural parameters in a given application. In practice,
however, subjective priors can be difficult to specify or controversial,
and it may be helpful to have default options. 

One common default is to use a flat prior, with $\pi$ proportional
to Lebesgue measure, where our assumption that $\Theta$ is compact ensures that this prior has finite mass.  As has previously been observed in other contexts, however, ``flatness'' of a prior is parameterization-specific, and the use of flat priors can lead two researchers, estimating the same model on the same data but with different parameterizations, to different posteriors. 
This motivates us to seek parameterization-invariant default priors.  Since the covariance function $\Sigma,$ and its domain $\Theta\times\Theta,$ are known in the limit experiment they can be used to inform such a prior.  Correspondingly, we define a default prior as a $\Sigma$-dependent probability measure on $\Theta$,  $\pi(\cdot;\Sigma)$. 

To formally define reparameterization-invariance, let $\Psi$ be a compact set and let $\vartheta:\Theta\to\Psi$ be a diffeomorphism between $\Theta$ and $\Psi,$ corresponding to a reparameterization $\psi=\vartheta(\theta)$ of the
model. This implies reparameterized moments
$
h\left(\cdot\right)=g\left(\vartheta^{-1}\left(\cdot\right)\right)
$
defined on $\Psi$, where by construction $h\left(\cdot\right)\sim\mathcal{GP}\left(m_{h},\Sigma_{h}\right)$
for 
\[
m_{h}\left(\psi\right)=m\left(\vartheta^{-1}\left(\psi\right)\right),\,\,\Sigma_{h}\left(\psi_{1},\psi_{2}\right)=\Sigma\left(\vartheta^{-1}\left(\psi_{1}\right),\vartheta^{-1}\left(\psi_{2}\right)\right).
\]
We call a rule for constructing a
prior invariant to re-parameterization if
the pushforward of $\pi\left(\cdot;\Sigma\right)$ under $\vartheta\left(\cdot\right)$
is equal to $\pi\left(\cdot;\Sigma_{h}\right)$ for all re-parametrizations $\vartheta\left(\cdot\right)$, so the default delivers the same prior (and thus posterior) distribution regardless of the parameterization.  The flat prior is invariant when $\vartheta$ is linear, but not generally otherwise.

The structure of the GMM model is also preserved under linear transformations of the moments.  Specifically, let $B:\Theta\to\mathcal{B}$ be a differentiable function
from $\Theta$ to the set $\mathcal{B}$ of full-rank $k\times k$
matrices. We can define a new moment process 
$
h\left(\theta\right)=B\left(\theta\right)g\left(\theta\right),
$ where by construction $h\left(\cdot\right)\sim\mathcal{GP}\left(m_{h},\Sigma_{h}\right)$
for 
\[
m_{h}\left(\theta\right)=B\left(\theta\right) m\left(\theta\right),\,\,\Sigma_{h}\left(\theta_{1},\theta_{2}\right)=B\left(\theta_{1}\right)\Sigma\left(\theta_{1}\,\theta_{2}\right)B\left(\theta_{2}\right)'.
\]
The moments $h$ and $g$ are one-to-one transformations of each other, and  imply the same value for the continuously updated GMM objective.  Hence, it is again natural to require that a default prior be invariant to such transformations, with $\pi\left(\cdot;\Sigma_{h}\right)=\pi\left(\cdot;\Sigma\right)$.

In parametric models the
desire for a parameterization-invariant default prior has led to
the use of the Jeffreys (1946) prior, which is usually defined to be proportional to $|i(\theta)|^{1/2}$, for $|i(\theta)|$ the determinant of the Fisher information.  Our suggested default prior is based on an analogous idea.  
Specifically, recall that the Fisher information is equal to the variance of the score, that is, the gradient of the log likelihood. We might analogously try to form a default prior based on the variance of the gradient of the GMM objective, $\operatorname{Var}(\frac{\partial}{\partial\theta}Q(\theta|g))$. Unfortunately, however, $\operatorname{Var}(\frac{\partial}{\partial\theta}Q(\theta|g))$ depends on the mean function $m,$ and so is unknown in general.  To construct a feasible default prior, we instead consider the variance of the continuously updated GMM objective in the fully-unidentified case, $i(\theta;\Sigma)=\operatorname{Var}(\frac{\partial}{\partial\theta}Q(\theta|G))$ for $G\sim \mathcal{GP}(0,\Sigma).$  The entries of $i(\theta;\Sigma)$ take the form
\begin{align}\label{eq: Jeffreys prior}
i_{jl}(\theta;\Sigma)=tr\left( \Sigma^{-1}(\theta,\theta)\left.\left\{\frac{\partial^2\Sigma(\theta,\tilde\theta)}{\partial\theta_j\partial\tilde\theta_l}-\frac{\partial \Sigma(\theta,\tilde\theta)}{\partial\theta_j}\cdot\Sigma^{-1}(\theta,\theta)\frac{\partial \Sigma(\tilde\theta,\theta)}{\partial\theta_l}\right\}\right|_{\tilde\theta=\theta}\right)
\end{align}
and so can be computed from $\Sigma$.
Since the continuously updated GMM objective is unchanged by linear transformations of the moments, $i_{jl}(\theta;\Sigma)$ is also unchanged.  Moreover, the same calculations which prove the invariance of Jeffreys prior for parametric models show that the default prior proportional to the square root of the determinant of $i(\theta;\Sigma),$
\begin{equation}\label{eq: default prior}
\pi(\cdot;\Sigma)\propto|i(\cdot;\Sigma)|^{1/2}
\end{equation}
is likewise invariant to reparameterization.  Hence, the default prior (\ref{eq: default prior}) is invariant to both reparameterization and linear transformations of the moments, as desired.\footnote{While we focus on the default prior (\ref{eq: default prior}), other invariant priors exist. For instance, for scalar $\theta$ the prior proportional to $
\left|\Sigma\left(\theta,\theta\right)\right|^{-\frac{1}{2}}\left|\frac{\partial^{2}}{\partial\theta\partial\tilde{\theta}}\Sigma\left(\theta,\tilde{\theta}\right)-\frac{\partial}{\partial\theta}\Sigma\left(\theta,\tilde{\theta}\right)\Sigma\left(\theta,\theta\right)^{-1}\frac{\partial}{\partial\tilde{\theta}}\Sigma\left(\theta,\tilde{\theta}\right)\right|_{\tilde{\theta}=\theta}^{\frac{1}{2}}
$
is also invariant.} 

While we motivated our default prior (\ref{eq: default prior}) by analogy to Jeffreys prior, it also has a more direct interpretation.  Since $G=g-m$ corresponds to the noise component of the GMM moments, the score $\frac{\partial}{\partial\theta}Q(\theta|G)$ measures the speed with which the noise in the moments changes at $\theta$.  One can show, however, that $E[\frac{\partial}{\partial\theta}Q(\cdot|G)]\equiv0$, so 
$\operatorname{Var}(\frac{\partial}{\partial\theta}Q(\theta|G))=E[\frac{\partial}{\partial\theta}Q(\theta|G)\frac{\partial}{\partial\theta}Q(\theta|G)']$ measures the average (squared) magnitude of the score. Hence, the default prior (\ref{eq: default prior}) assigns more mass to regions of the parameter space where the noise component of the moments tends to change quickly in $\theta$, and less to regions where the noise tends to change slowly.
\paragraph{Special Case: Linear IV (continued)}
To explore the implications of our default prior in this example, let us partition the $2k\times 2k$ variance matrix into four $k\times k$ sub-matrices $\Omega_{jl}$ for $j,l\in\{0,1\},$ where $\Omega_{jl}=\operatorname{Cov}(\xi_j,\xi_l)$. Under this notation $i\left(\theta;\Sigma\right)$ is equal to the trace of
\[
\begin{array}{c}
\left(\Omega_{00}-\left(\Omega_{10}+\Omega_{01}\right)\theta+\Omega_{11}\theta^{2}\right)^{-1 }\times\\
\left(\Omega_{11}-\left(\Omega_{01}-\Omega_{11}\theta\right)\left(\Omega_{00}-\left(\Omega_{10}+\Omega_{01}\right)\theta+\Omega_{11}\theta^{2}\right)^{-1}\left(\Omega_{10}-\Omega_{11}\theta\right)\right),
\end{array}
\]
which can also be written as the relative variance:
\[
\operatorname{Var}\left(g\left(\theta\right)\right)^{-1} \operatorname{Var}\left(\xi_{1}|g\left(\theta\right)\right).
\]
Thus, the default prior (\ref{eq: default prior}) favors parameter values where (i) the GMM moments predict the first-stage poorly (so $\operatorname{Var}\left(\xi_{1}|g\left(\theta\right)\right)$ is large) and (ii) the GMM moments themselves are not too noisy (so $\operatorname{Var}\left(g\left(\theta\right)\right)$ is small). Since $\operatorname{Var}\left(\xi_{1}|g\left(\theta\right)\right)\to 0$  and $\operatorname{Var}\left(g\left(\theta\right)\right)\to\infty$ as $|\theta|\to\infty,$ we see that the default prior density (\ref{eq: default prior}) converges to zero for $\theta$ large.

We obtain further simplifications when the reduced-form and first-stage errors are homoskedastic.  In this case the matrices $\Omega_{jl}$ are proportional to each other,
$$
\Omega_\xi=\left(\begin{array}{cc}
   \sigma_u^2 & \sigma_{uv} \\
    \sigma_{uv} & \sigma_v^2
\end{array}\right)\otimes\widetilde\Omega,
$$
for $\widetilde\Omega$ a $k\times k$ matrix, and  $\sigma_u^2,$ $\sigma_v^2,$ $\sigma_{uv}$ again the variances and covariance of the reduced-form and first-stage errors.  In this case some algebra shows that our default prior is equal to a Cauchy distribution centered at  $\frac{\sigma_{uv}}{\sigma_v^2}$.  One can show, however, that $\frac{\sigma_{uv}}{\sigma_v^2}$ corresponds to the probability limit of the OLS estimator under weak-instrument asymptotics.  Hence in the homoskedastic case our default prior corresponds to a Cauchy distribution centered at OLS. $\Box$

\section{Linear IV Simulations}\label{sec: simulations}

While our theoretical results show that GMM estimators are dominated
under bounds on identification strength, they do not imply that GMM
is dominated by either bagged GMM or quasi-Bayes. Relative performance
of these estimators in applications is thus an open question. 
We explore this comparison in the context of linear IV, using
simulation designs based on Andrews et al. (2019).
Andrews et al. (2019) calibrated simulations based on all instrumental variables specifications published in the American Economic Review from 2014 to 2018 for which sufficient information is available to estimate
the variance matrix $\Omega_\xi$ in (\ref{eq: IV Model}), yielding 124 specifications.  We follow their simulation designs, and draw data from the normal model (\ref{eq: IV Model}) with $\pi^{*}$ equal to the first-stage estimate in the Andrews et al. (2019) data and $\theta^{*}$ equal to the two-stage least squares estimate.
We consider six different estimators.  The first two are GMM, specifically
two-stage least squares, which as noted above corresponds to
GMM with weighting matrix $\Xi\left(\theta\right)=E\left[Z_{i}Z_{i}'\right]^{-1}$, and
continuously updated GMM, which corresponds to GMM with weighting
matrix $\Xi\left(\theta\right)=\Sigma\left(\theta\right)^{-1}$. We next report bagged versions of each GMM estimator.  Finally, we report two quasi-Bayes estimators, the first using a flat prior $\pi\left(\theta\right)\propto1$ and the second using the invariant default prior (\ref{eq: default prior}).

Our theoretical results require the loss function to be bounded.  In the context of the linear IV model, this means that our results to not apply to unconstrained estimation of $\theta$ under squared error loss.  We address this in two ways, first considering estimation of $\theta$ under bounds on the parameter space and then considering estimation of the correlation between the structural and first-stage errors, which is a naturally bounded quantity.

In most economic applications of IV methods there are context-specific bounds on the range of plausible parameter values, for instance bounds beyond which a researcher would conclude that there is a problem with their data or their instrument.  Motivated by this observation, we take the parameter space in specification
$s$ to be $\Theta_{s}=\left[\pm 20 \left|\frac{\sigma_{uv,s}}{\sigma_{v,s}^2}\right| \right]$ for $\sigma_{uv,s}$ the covariance of the reduced-form and first-stage errors in specification $s$, and $\sigma_{v,s}^2$ the variance of the first-stage error. 
As noted above $\frac{\sigma_{uv}}{\sigma_v^2}$ corresponds to the probability limit of the least squares estimator under weak instrument asymptotics, so this parameter space restricts the size IV coefficient to be no more than 20 times the size of the least squares coefficient.  Nine of the just-identified specifications in the Andrews et al. (2019) data do not report estimates for $(\sigma_{uv},\sigma_v^2)$, because replication data were not publicly available but $\Omega_\xi$ could be estimated based on published results.  In these cases we set $\sigma_{uv}$ equal to covariance of the reduced-form and first-stage estimates, and $\sigma_v^2$ equal to the variance of the first-stage estimate.  Consistent with our motivation for restricted the parameter space, $\Theta_s$ contains the true IV coefficient in all but three of the 124 Andrews et al. (2019) specifications, so we limit attention to these 121 specifications for our analysis. We approximate the bagged estimators by averages over 400 bootstrap draws, and report results based on 10,000 simulation draws.

The restriction to a bounded parameter space is substantively important for the results we report below. Absent a restriction on the parameter space, the two-stage least squares estimator for the IV coefficient only has finite moments up to the degree of over-identification.  Correspondingly, for an unrestricted parameter space the bagged two-stage least squares estimate is only defined for $k\ge2$.  The situation is, if anything, worse for the other estimators we consider: for instance, in linear IV models with homoskedastic errors the continuously updating GMM estimator reduces to the limited information maximum likelihood estimator, which lacks even a first moment.  Our restriction to a bounded parameter space ensures that all estimators and moments are well-defined and finite, but for those cases where the restriction is important for e.g. existence of a given moment, the results necessarily depend on the choice of bounds.

Motivated by this sensitivity, we report three additional sets of results.  In the main text we report results estimating the correlation between the structural and first-stage errors, which can be shown to equal $r(\theta)=\frac{\sigma_{uv}-\theta\sigma_v^2}{\sigma_v\sqrt{\sigma_u^2-2\theta\sigma_{uv}+\theta^2\sigma_v^2}}.$  This correlation measures the degree of endogeneity, and so may be of interest in its own right.  Moreover, as recently highlighted by Angrist and  Kolesár (2022), conventional confidence intervals are reliable in just-identified IV settings, even with weak instruments, so long as this correlation coefficient is not too large.  Hence, we might also want to know $r(\theta)$ for that reason. Importantly for our current purposes, $r(\theta)$ is bounded by construction.

In the appendix we report results where we vary the definition of the parameter space, considering $\Theta_{s}=\left[\pm 40 \left|\frac{\sigma_{uv,s}}{\sigma_{v,s}^2}\right| \right]$ and $\Theta_{s}=\left[\pm 60 \left|\frac{\sigma_{uv,s}}{\sigma_{v,s}^2}\right| \right].$  Widening the bounds increases the errors of the IV coefficient estimates from all approaches, but the ordering of estimators by average performance is unchanged.  As expected, the change in the bounds has almost no effect on the correlation coefficient estimates.  Finally we report results, for two-stage least squares and bagged two-stage least squares only, which drop the bounds on the parameter space entirely.  To ensure that the mean squared error is well-defined these results restrict attention to specifications with at least three instruments, $k\ge 3$.  The performance gaps between these estimators are small in our $k\ge3$ specifications with bounded parameter spaces, and remain small when we drop the bounds.

\paragraph{Results for IV Coefficient}

\begin{figure}
\includegraphics[scale=0.6]{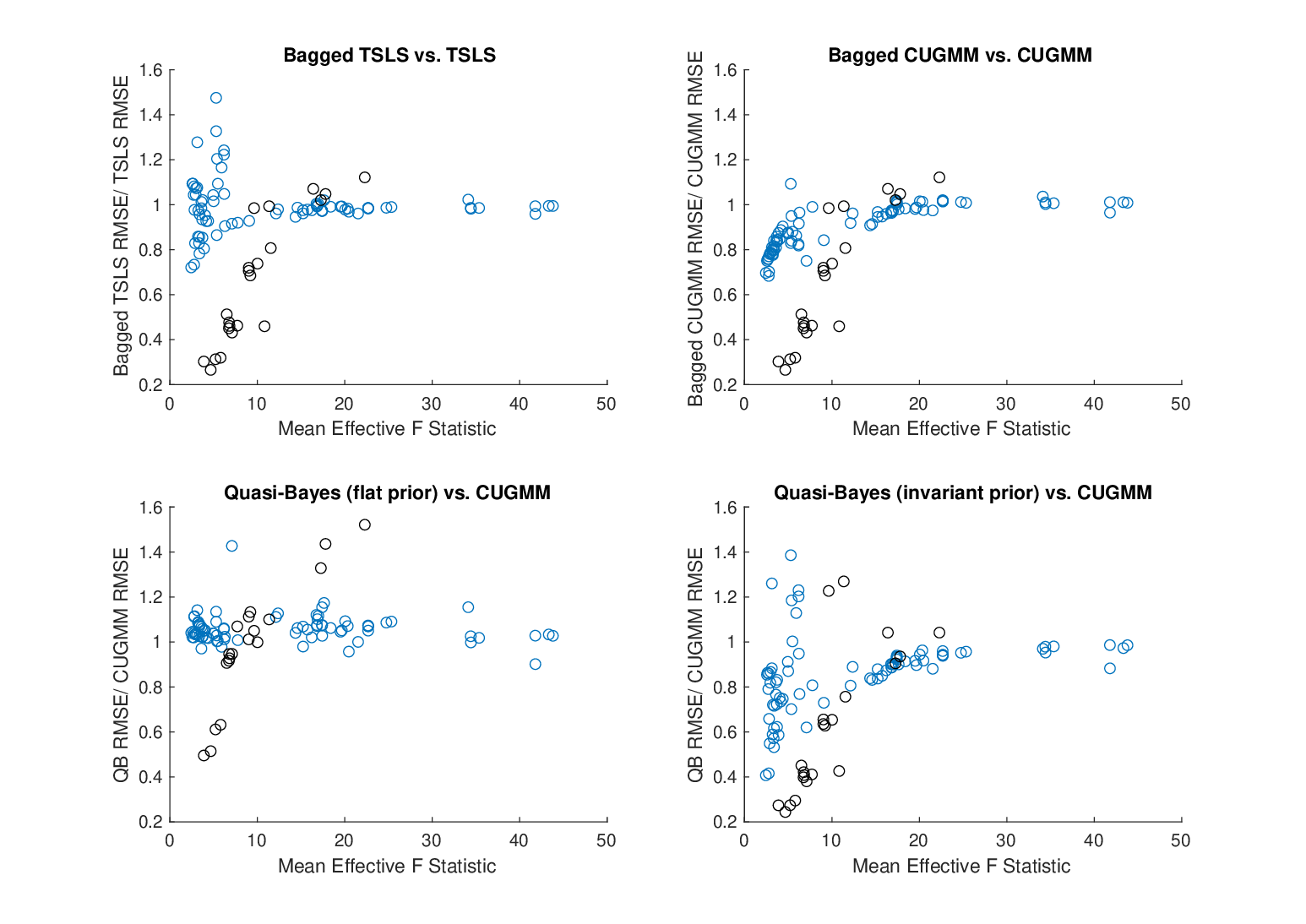}
\caption{RMSE comparisons for IV coefficient estimators in Andrews et al. (2019) specifications. Each point corresponds to one of the Andrews et al. (2019) specifications. The vertical axis measures the ratio of root mean squared error for the alternative estimator compared to the GMM estiamtor for the IV coefficient, $\sqrt{\mathbb{E}_s[(\delta(g)-\theta^*_s)^2]/\mathbb{E}_s[(\delta^{GMM}(g)-\theta^*_s)^2]}$.  So, for instance, a value of 0.8 means the RMSE for the alternative estimator is 20\% lower.  The horizontal axis shows the average effective first-stage F statistic of Montiel-Olea and Pflueger (2013), $\mathbb{E}_s[F]$.   We limit attention to specifications with $\mathbb{E}_s[F]\le 50$ for visibility.  Blue dots correspond to just-identified ($k=1$) specifications, while black dots correspond to over-identified ($k\ge 2$) specifications.\label{fig: simulation results}}
\end{figure}

Table \ref{tab: Simulation Results} reports our findings for the IV coefficient. For each
specification $s$ we consider the root mean squared error for each
estimator, normalized by the two-stage least squares standard error $\sigma_s^*$ in the Andrews et al. (2019) data to account for differences in units, $\sqrt{\mathbb{E}_{s}\left[\left(\delta\left(g\right)-\theta_{s}^{*}\right)^{2}\right]}/\sigma_{s}^{*},$
where $\mathbb{E}_{s}\left[\cdot\right]$ denotes the expectation
in specification $s$. We report the average of this ratio for each
estimator across four different categories based on the effective
first-stage F statistic of Montiel-Olea and Pflueger (2013). The effective
F statistic, which we denote by $F$, is a measure of instrument strength
and in the just-identified case is equal to the squared t-statistic for testing $\pi^{*}=0$, $\xi_{1}^{2}/\operatorname{Var}\left(\xi_{1}\right)$.
See Montiel-Olea and Pflueger (2013) for details and motivation for this
statistic.  To complement these results, Figure \ref{fig: simulation results} plots the root mean squared error for each alternative estimator, relative to its GMM counterpart, against the average effective F statistic $\mathbb{E}_s[F]$, limiting attention to specifications where $\mathbb{E}_s[F]\le 50$ for visibility.  To show differences based on the number of instruments we plot just-identified ($k=1$) specifications in blue, and over-identified ($k\ge2$) specifications in black.  For a more detailed picture of how performance varies with the number of instruments, Table \ref{tab: Simulation Results, k} in the appendix reports average results when we bin specifications by the number of instruments.

A number of patterns emerge in Table \ref{tab: Simulation Results} and Figure \ref{fig: simulation results}.
First, two-stage least squares outperforms continuously updating GMM
everywhere except the strongest identification category. The bagged GMM estimators each outperform their standard GMM
analogs in most cases. Specifically, these estimators show substantial
improvements in the category where identification is weakest ($\mathbb{E}_{s}\left[F\right]\le10$),
a smaller improvement in the next-weakest category ($10<\mathbb{E}_{s}\left[F\right]\le20$),
and either a minimal improvement or a small deterioration in the second-strongest category
($20<\mathbb{E}_{s}\left[F\right]\le50$).  Quasi-Bayes with a flat prior underperforms all the other estimators, while quasi-Bayes with the invariant prior outperforms both GMM estimators except in the second-strongest category ($20<\mathbb{E}_{s}\left[F\right]\le50$). The performance gap between the two quasi-Bayes approaches demonstrates the influence of the prior, and highlights that the greater smoothness of quasi-Bayes as a function of the moments does not guarantee improved performance for all priors.
Finally, all estimators show very similar performance in the strongest category ($50<\mathbb{E}_{s}\left[F\right]$), though two-stage least squares and bagged two-stage least squares are known to be inefficient under strong identification. Nonetheless, the estimator with the best average performance overall is bagged two-stage least squares, followed by bagged continuously updating GMM.

\begin{table}
\begin{tabular}{ccccc}
 & $\mathbb{E}_{s}\left[F\right]\le10$  & $10<\mathbb{E}_{s}\left[F\right]\le20$  & $20<\mathbb{E}_{s}\left[F\right]\le50$  & $50<\mathbb{E}_{s}\left[F\right]$\tabularnewline
\hline 
\hline 
$\begin{array}{c}
\text{Two-Stage}\\
\text{ Least Squares}
\end{array}$  & 1.37 & 1.20 & 1.02 & 1.00\tabularnewline
\hline 
$\begin{array}{c}
\text{Continuously}\\
\text{ Updating GMM}
\end{array}$  & 1.55 & 1.24 & 1.03 & 0.99\tabularnewline
\hline 
$\begin{array}{c}
\text{Bagged Two-Stage}\\
\text{ Least Squares}
\end{array}$  & 1.03 & 1.09 & 1.01 & 0.99\tabularnewline
\hline 
$\begin{array}{c}
\text{Bagged Continuously}\\
\text{ Updating GMM}
\end{array}$  & 1.04 & 1.12 & 1.05 & 0.99\tabularnewline
\hline 
$\begin{array}{c}
\text{Quasi-Bayes,}\\
\text{ Flat Prior}
\end{array}$  & 1.51 & 1.61 & 1.10 & 1.00\tabularnewline
\hline 
$\begin{array}{c}
\text{Quasi-Bayes,}\\
\text{Invariant Prior}
\end{array}$  & 1.18 & 1.10 & 1.03 & 0.99\tabularnewline
\hline 
$\text{Number of Specifications}$  & 56 & 28 & 19 & 18\tabularnewline
\end{tabular}

\caption{Performance of IV coefficient estimators in Andrews et al. (2019) specifications. Entries
correspond to the root mean squared error normalized by the standard
error in the Andrews et al. (2019) data, $\sqrt{\mathbb{E}_{s}[(\delta(g)-\theta_{s}^{*})^{2}]}/\sigma_{s}^{*}$,
averaged across specifications. Columns correspond to ranges of values
for the average effective first-stage F statistic of Montiel-Olea
and Pflueger (2013).\label{tab: Simulation Results}}
\end{table}

\paragraph{Results for Correlation Coefficient}

Table \ref{tab: Simulation Results, corr} reports our findings for the correlation coefficient $r(\theta)=\frac{\sigma_{uv}-\theta\sigma_v^2}{\sigma_v\sqrt{\sigma_u^2-2\theta\sigma_{uv}+\theta^2\sigma_v^2}}.$
We again report the root mean squared error normalized by the delta-method standard error $\sigma_{r,s}^*$ for $r(\theta)$ in the Andrews et al. (2019) data, $\sqrt{\mathbb{E}_{s}\left[\left(\delta\left(g\right)-r(\theta_{s}^{*})\right)^{2}\right]}/\sigma_{r,s}^{*},$ and bin specifications based on the average effective first-stage F statistic.  Figure \ref{fig: simulation results, corr} plots the root mean squared error for each alternative estimator, relative to its GMM counterpart, against the average effective F statistic $\mathbb{E}_s[F]$.  Table \ref{tab: Simulation Results, k, corr} in the appendix reports average results when we bin specifications by the number of instruments. 

The results for estimating the correlation coefficient are broadly consistent with those for the IV coefficient.  In particular, two-stage least squares largely outperforms continuously updating GMM, and each bagged estimator substantially outperforms its GMM counterpart in the weakest category, with smaller gain in the stronger categories.  Quasi-Bayes with a flat prior again underperforms the other estimators considered.  One difference with the IV coefficient results is that quasi-Bayes with our invariant prior now slightly under-performs relative to two-stage least squares.  All estimators again behave very similarly in the strongest specifications ($50<\mathbb{E}_{s}\left[F\right]$).  The best-performing estimator on average is bagged continuously updating GMM, followed by bagged two-stage least squares.

Overall, these simulation results show that bagged GMM estimators often  outperform their standard GMM counterparts in the settings we consider, through neither dominates the other.  Quasi-Bayes with a flat prior performs quite poorly, while quasi-Bayes with the invariant prior (\ref{eq: default prior}) is more competitive.

\begin{table}
\begin{tabular}{ccccc}
 & $\mathbb{E}_{s}\left[F\right]\le10$  & $10<\mathbb{E}_{s}\left[F\right]\le20$  & $20<\mathbb{E}_{s}\left[F\right]\le50$  & $50<\mathbb{E}_{s}\left[F\right]$\tabularnewline
\hline 
\hline 
$\begin{array}{c}
\text{Two-Stage}\\
\text{ Least Squares}
\end{array}$  & 1.09 & 1.10 & 1.01 & 0.99\tabularnewline
\hline 
$\begin{array}{c}
\text{Continuously}\\
\text{ Updating GMM}
\end{array}$  & 1.21 & 1.13 & 1.02 & 0.99\tabularnewline
\hline 
$\begin{array}{c}
\text{Bagged Two-Stage}\\
\text{ Least Squares}
\end{array}$  & 1.03 & 1.07 & 1.00 & 0.99\tabularnewline
\hline 
$\begin{array}{c}
\text{Bagged Continuously}\\
\text{ Updating GMM}
\end{array}$  & 0.99 & 1.09 & 1.02 & 0.99\tabularnewline
\hline 
$\begin{array}{c}
\text{Quasi-Bayes,}\\
\text{ Flat Prior}
\end{array}$  & 1.19 & 1.45 & 1.07 & 0.99\tabularnewline
\hline 
$\begin{array}{c}
\text{Quasi-Bayes,}\\
\text{Invariant Prior}
\end{array}$  & 1.10 & 1.11 & 1.01 & 0.99\tabularnewline
\hline 
$\text{Number of Specifications}$  & 56 & 28 & 19 & 18\tabularnewline
\end{tabular}
\caption{Performance of correlation coefficient estimators in Andrews et al. (2019) specifications. Entries
correspond to the root mean squared error normalized by the standard
error in the Andrews et al. (2019) data, $\sqrt{\mathbb{E}_{s}[(\delta(g)-r(\theta_{s}^{*}))^{2}]}/\sigma_{r,s}^{*}$,
averaged across specifications. Columns correspond to ranges of values
for the average effective first-stage F statistic of Montiel-Olea
and Pflueger (2013).\label{tab: Simulation Results, corr}}
\end{table}

\begin{figure}
\includegraphics[scale=0.6]{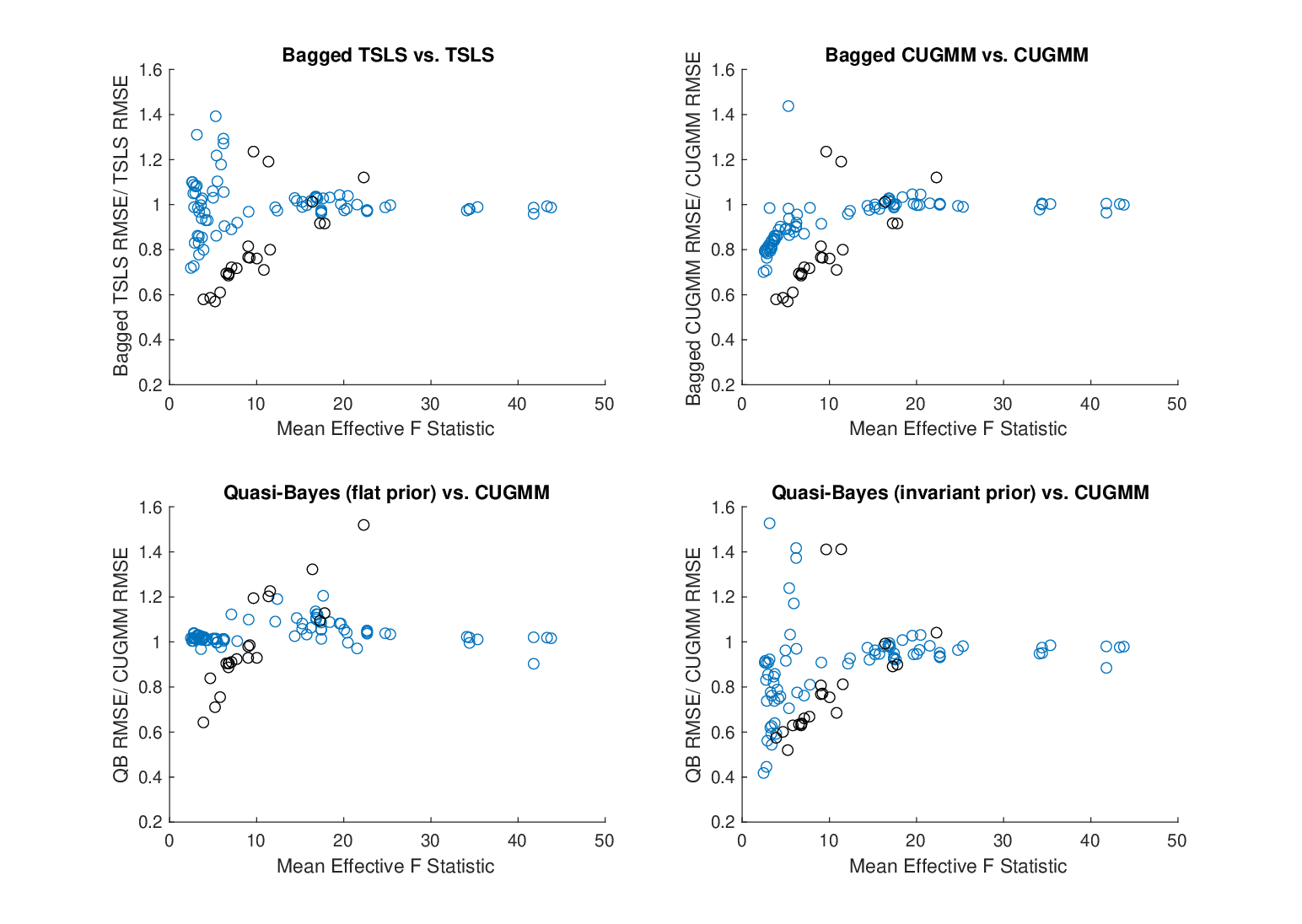}
\caption{RMSE comparisons for correlation coefficient estimators in Andrews et al. (2019) specifications. Each point corresponds to one of the Andrews et al. (2019) specifications. The vertical axis measures the ratio of root mean squared error for the alternative estimator compared to the GMM estiamtor for the correlation coefficient, $\sqrt{\mathbb{E}_s[(\delta(g)-r(\theta^*_s))^2]/\mathbb{E}_s[(\delta^{GMM}(g)-r(\theta^*_s))^2]}$.  So, for instance, a value of 0.8 means the RMSE for the alternative estimator is 20\% lower.  The horizontal axis shows the average effective first-stage F statistic of Montiel-Olea and Pflueger (2013), $\mathbb{E}_s[F]$.   We limit attention to specifications with $\mathbb{E}_s[F]\le 50$ for visibility.  Blue dots correspond to just-identified ($k=1$) specifications, while black dots correspond to over-identified ($k\ge 2$) specifications.\label{fig: simulation results, corr}}
\end{figure}

\clearpage

\section{References}

\noindent Adler, R.J., and J.E. Taylor (2007): \emph{Random Fields and Geometry}, New York : Springer

\noindent  Andrews, D.W.K. and X. Cheng (2012):
``Estimation and Inference with Weak, Semi-strong and Strong
Identification,'' \emph{Econometrica}, 80(5), 2153-2211.

\noindent  Andrews, I. (2019): ``On the Structure of IV Estimands,'' \emph{Journal of Econometrics},  211(1), 294-307.

\noindent  Andrews, I. and A. Mikusheva (2022): ``Optimal Decision Rules for Weak GMM,'' \emph{Econometrica}, 90(2), 715-748.

\noindent Andrews, I., J. Stock, and L. Sun, (2019): ``Weak Instruments in IV Regression: Theory and Practice.'' \emph{Annual Review of Economics}, 11, 727-753.

\noindent Angrist, J. and M. Kolesár (2022): ``One Instrument to Rule Them All: The Bias and Coverage of Just-ID IV,'' Working Paper, https://arxiv.org/abs/2110.10556.

\noindent Armstrong, T. (2016): ``Large Market Asymptotics for Differentiated Product Demand Estimators With Economic Models of Supply,'' \emph{Econometrica}, 84(5), 1961-1980.

\noindent Armstrong, T. and M. Kolesar (2021): ``Sensitivity Analysis Using Approximate Moment Condition Models,'' \emph{Quantitative Economics}, 12(1), 77-108.

\noindent Berlinet, A. and C. Thomas-Agnan (2004): ``Reproducing Kernel Hilbert Spaces in Probability and Statistics,'' Spinger.

\noindent Bühlmann, P. and B. Yu (2002): ``Analyzing Bagging,'' \emph{The Annals of Statistics}, 30(4), 927-961.

\noindent Brown, L.D.  (1986): \emph{Fundamentals of Statistical Exponential Families with Applications in Statistical Decision Theory}, Hayward, CA: Institute of Mathematical Statistics.

\noindent Chamberlain, G. (1987): ``Asymptotic Efficiency in Estimation With Conditional Moment Restrictions,'' \emph{Journal of Econometrics}, 34(3), 305-334.

\noindent  Chen, S.X. and P. Hall (2003): ``Effects of Bagging and Bias Correction on Estimators Defined by Estimating Equations,'' \emph{Statistica Sinica}, 13(1), 97-109.

\noindent  Chernozhukov, V. and H. Hong (2003): ``An MCMC Approach to Classical Estimation,'' \emph{Journal of Econometrics}, 115(2), 293-346.

\noindent Guggenberger, P. and R. Smith (2005): ``Generalized Empirical Likelihood Estimators and Tests Under
Partial, Weak, and Strong Identification,'' \emph{Econometric Theory}, 21 (4), 667–709.

\noindent Hansen, L.P. (1982) : ``Large Sample Properties of Generalized Method of Moments Estimators,'' \emph{Econometrica}, 50(4), 1029-1054.

\noindent Horowitz, J.L. (2001): ``The Bootstrap,'' Handbook of Econometrics, Volume 5, Eds. James J. Heckman, Edward Leamer, 3159-3228.

\noindent Huber, P.J. (1964): ``Robust Estimation of a Location Parameter,'' \emph{The Annals of Mathematical Statistics}, 35(1), 73-101.

\noindent Jeffreys, H. (1946): ``An Invariant Form of the Prior in Estimation Problems,'' \emph{Proceedings of the Royal Society of London. Series A, Mathematical and Physical Sciences}, 186(1007), 453-461.

\noindent Manski, C.F. (2021) : ``Econometrics for Decision Making: Building Foundations Sketched by Haavelmo and Wald,'' \emph{Econometrica}, 89(6), 2827-2853.

\noindent Mavroeidis, S., M. Plagborg-Møller, and J.H. Stock (2014): ``Empirical Evidence on Inflation Expectations in the New Keynesian Phillips Curve.'' \emph{Journal of Economic Literature} 52(1): 124-188.

\noindent Montiel-Olea J. and C. Pflueger C (2013): ``A Robust Test for Weak Instruments.'' \emph{Journal of Business and Economic Statistics} 31(3): 358-369.

\noindent Staiger, D. and J.H. Stock (1997): ``Instrumental Variables Regression with Weak Instruments,'' \emph{Econometrica}, 65(3), 557-586.

\noindent Stock, J.H. and J. Wright (2000): ``GMM with Weak Identification,''
\emph{Econometrica}, 68, 1055-96.

\noindent  Van der Vaart, A.W. and J.A. Wellner (1996):  \emph{Weak Convergence and Empirical Processes}, Springer.

\noindent  Van der Vaart, A.W. and H. Van Zanten (2008): ``Reproducing Kernel Hilbert Spaces of Gaussian Processes,'' in \emph{Pushing the Limits of Contemporary Statistics: Contributions in Honor of Jayanta K. Ghosh},  Bertrand Clarke and Subhashis Ghosal, eds., (Beachwood, Ohio, USA: Institute of Mathematical Statistics, 2008), 200-222.

\begin{appendix}

\section{Proofs}

\paragraph{Proof of Lemma \ref{lem: Pettis Representation}}

If $k=1$,  the result is immediate from Theorem 2.1 of van der Vaart and van Zanten (2008).  We are left to prove it for $k>1$.
Define an augmented parameter space $\Theta^*=\Theta\times V,$ where  $V=\{v\in\mathbb{R}^k:\|v\|_1=1\}$ and consider a Gaussian process $g^*(\cdot)$ defined on $\Theta^*$ as $g^*(\theta,v)=v'g(\theta).$   Note that the process $g^*$, its mean  $m^*(\theta,v)$ and its covariance function $\Sigma^*(\theta,v,\tilde\theta,\tilde v)$ are one-to-one transformations of $g,$ $m,$ and $\Sigma.$ For $\mathcal{H}^*$ the RKHS associated with $\Sigma^*,$ $\mathcal{H}^*$ is isometric to $\mathcal{H}$. Indeed, for $m^*\in\mathcal{H}^*$:
\begin{align}\label{eq: connection between RKHS}
m^*(\theta,v)=\sum\alpha_i\Sigma^*(\theta_i,v_i,\theta,v)=\left(\sum\alpha_iv_i'\Sigma(\theta_i,\theta)\right)v=m(\theta)'v,
\end{align}
where $m\in\mathcal{H}$ and $\|m\|_{\mathcal{H}}=\|m^*\|_{\mathcal{H}^*}$.

We have assumed that $g(\cdot)\sim \mathcal{GP}(m,\Sigma)$ has almost-surely continuous sample paths, which implies that $G^*\sim \mathcal{GP}(0,\Sigma^*)$ can likewise be realized as a process with almost surely continuous sample paths. Let $\mathcal{C}^*$ be the space of $\mathbb{R}$-valued continuous functions on $\Theta^*$ with the property that any $f^*\in \mathcal{C}^*$ can be represented as $f^*(\theta,v)=f(\theta)'v$ for $f\in\mathcal{C}(\Theta,\mathbb{R}^k)$ and $v\in V$. Due to the structure of  $\Sigma^*,$ realizations of the process $G^*$ almost surely belong to $\mathcal{C}^*$ and the process can be represented as $G^*(\theta,v)=v'G(\theta)$, where $G\sim \mathcal{GP}(0,\Sigma).$  Take any linear functional defined on the space of continuous functions with index set $\Theta^*$ and denote by $\eta^*$ its restriction to $\mathcal{C}^*$. Since the  relation between $f^*\in\mathcal{C}^*$ and $f\in\mathcal{C}(\Theta,\mathbb{R}^k)$ is one-to-one, we can define a linear functional on $\mathcal{C}(\Theta,\mathbb{R}^k)$ as $\eta(f)=\eta^*(f^*)$. This creates a one-to-one correspondence between linear functionals on $\mathcal{C}^*$ and linear functionals on $\mathcal{C}(\Theta,\mathbb{R}^k)$. Note that the definition of the Pettis integral for process $G^*$ depends on $\eta^*$ only and all functionals that are the same once restricted to $\mathcal{C}^*$ lead to the same Pettis integral: 
\begin{align}\label{eq: relations between ms}
m^*_{\eta^*}(\theta,v)=\mathbb{E}[G^*(\theta,v)\eta^*(G^*)]=\mathbb{E}[v'G(\theta)\eta^*(G^*)]=v'\mathbb{E}[G(\theta)\eta(G)]=v'm_{\eta}(\theta).
\end{align}
Due to Theorem 2.1 of van der Vaart and van Zanten (2008), $\mathcal{H}^*$ coincides with the image of  the space of linear functionals defined on $\mathcal{C}^*$ under the Pettis integral transformation, while equation (2.4) in that paper together with the definition of $\sigma(G^*)$ establish that $\|m_{\eta^*}\|_{\infty}\leq\sigma^2(G^*)\|\eta^*\|^*$. Comparing  equation (\ref{eq: relations between ms}) to (\ref{eq: connection between RKHS}), we see that the first statement of Lemma \ref{lem: Pettis Representation} holds. 
We further notice that all norms of starred objects coincide with the norms of the corresponding objects without stars. For example,  
\[
\|m\|_\infty=\sup_{j,\theta\in\Theta}|m_j(\theta)|=
\sup_{v\in V,\theta\in\Theta}|v'm(\theta)|=\sup_{(\theta,v)\in\Theta^*}|m^*(\theta,v)|=\|m^*\|_\infty.
\]
Note further that $\eta$ and $\eta^*$ have the same total variation norm.
$$
\|\eta^*\|^*=\sup_{f^*\in\mathcal{C}^*,\|f^*\|_\infty\le1}\eta^*(f^*)=\sup_{f\in\mathcal{C}(\Theta,\mathbb{R}^k),\|f\|_\infty\le1}\eta(f)=\|\eta\|^*.
$$
Finally, 
$$\sigma^2(G^*)=\sup_{\|\eta^*\|^*\le 1}\mathbb{E}[\eta^*(G^*)^2]=\sup_{\|\eta\|^*\le 1}\mathbb{E}[\eta(G)^2]=\sigma^2(G).
$$
This completes the proof. $\Box$

\begin{theorem}\label{Thm: Brown admissibility} (Brown 1986, Andrews and Mikusheva 2022): For any parameter space $\widetilde{\Gamma}\subseteq \Gamma$, any loss $L(a,\theta)$ which is convex in $a$ for all $\theta$, and any decision rule $\delta$ that is admissible on $\widetilde{\Gamma}$, there exists a sequence of finitely supported priors $\pi_r$ on $\widetilde{\Gamma}$ and corresponding Bayes decision rules $\delta_{\pi_r}$,
$$
\int \mathbb{E}_{m}[L(\delta_{\pi_r}(g),\theta^*)]d\pi_r(\theta^*,m)=\min_{\tilde{\delta}} \int \mathbb{E}_{m}[L(\tilde{\delta}(g),\theta^*)]d\pi_r(\theta^*,m),
$$
such that $\delta_{\pi_r}(g)\to \delta(g)$ as $r\to \infty$ for almost every $g$.
\end{theorem}

\paragraph{Proof of Theorem \ref{thm: ae continuity}}
Consider the posterior risk function 
$
R_{\pi}\left(a,g\right)=\mathbb{E}_{\pi}\left[L\left(a,\theta\right)|g\right].
$
As the first step of the proof we show that for any  finitely-supported
prior $\pi$ on $\widetilde{\Gamma}$, the risk function $R_{\pi}\left(a,g\right)$
is Lipschitz in $g$:
\begin{align}\label{eq: risk is lipshitz}
    \left| R_{\pi}\left(a,g\right)-R_{\pi}\left(a,g'\right)\right| \le \bar{L}W\left\| g-g'\right\|_\infty.
\end{align}
Let $\left\{ \left(\theta_{1},m_{1}\right),...,\left(\theta_{J},m_{J}\right)\right\} $
be the support of $\pi$. 
For each $m_j$ we know  from Lemma \ref{lem: Pettis Representation} that there exists $\eta_{m,j}\in\mathbb{H}$ with $\|\eta\|^*\le W$ and $m_j(\cdot)=\mathbb{E}[G(\cdot)\eta_{m,j}(G)].$ Further note that by e.g. Lemma 3.1 of van der Vaart and van Zanten (2008), for each $m\in\mathcal{H}$ the likelihood ratio for the measure $Q_m$ corresponding to a $\mathcal{GP}(m,\Sigma)$ distribution, relative to $m'=0$, takes the form $
\frac{dQ_m}{dQ_0}(g)=\exp\left(\eta_m(g)-\frac{1}{2}\|m\|_{\mathcal{H}}^2\right).
$

Define
$\tilde{g}=g'-g$, and let $g_{t}=g+t\cdot\tilde{g}$. 
Note that
\begin{align}\label{eq: conditional expectation}
 \mathbb{E}_{\pi}\left[L\left(a,\theta\right)|g_t\right]=\frac{\sum_{j}L\left(a,\theta_j\right)\exp\left\{ \eta_{m,j}(g_t) -\frac{1}{2}\left\Vert m_{j}\right\Vert _{\mathcal{H}}^{2}\right\} \pi\left(\theta_{j},m_{j}\right)}{\sum_{j}\exp\left\{ \eta_{m,j}(g_t)  -\frac{1}{2}\left\Vert m_{j}\right\Vert _{\mathcal{H}}^{2}\right\} \pi\left(\theta_{j},m_{j}\right)}.
\end{align}
Linearity implies that $\eta_{m,j}(g_t)=\eta_{m,j}(g)+t\eta_{m,j}(\tilde g),$ and thus
\[
\frac{\partial}{\partial t}\exp\left\{ \eta_{m,j}(g_t) -\frac{1}{2}\left\Vert m_{j}\right\Vert _{\mathcal{H}}^{2}\right\} =\eta_{m,j}(\tilde g)\exp\left\{ \eta_{m,j}(g_t) -\frac{1}{2}\left\Vert m_{j}\right\Vert _{\mathcal{H}}^{2}\right\} .
\]
By differentiating (\ref{eq: conditional expectation}) we get 
\begin{equation}\label{eq: risk derivative}
\frac{\partial}{\partial t}\mathbb{E}_{\pi}\left[L\left(a,\theta\right)|g_t\right]=\operatorname{Cov}_{\pi}\left(L\left(a,\theta\right),\eta_{m}(\tilde g) |g_t\right),
\end{equation}
where the only posterior uncertainty about $\eta_{m}(\tilde g)$ comes from the unknown parameter $m$, while $\tilde g$ is fixed. The Cauchy-Schwarz inequality implies that
\[
\left\Vert \operatorname{Cov}_{\pi}\left(L\left(a,\theta\right),\eta_{m}(\tilde g)|g_t\right)\right\Vert \le \bar{L}\sqrt{\operatorname{Var}(\eta_{m}(\tilde g)|g_t)}.
\]
For $(\theta_j,m_j)\in\Gamma_W$ we have $\|\eta_{m,j}\|^*\leq W$, thus
$|\eta_{m,j}(\tilde g)|\le W\|\tilde g\|_\infty= W\|g-g'\|_\infty,$ 
which implies that
$\sqrt{\operatorname{Var}(\eta_{m}(\tilde g)|g_t)}\le W\|g-g'\|_\infty.$
Hence,
\[
\left|R_{\pi}\left(a,g\right)-R_{\pi}\left(a,g'\right)\right|=
\left|\int_{0}^{1}\frac{\partial}{\partial t}\mathbb{E}_{\pi}\left[L\left(a,\theta\right)|g_{t}\right]dt\right|\le \bar{L}W\left\Vert g-g'\right\Vert _{\mathcal{\infty}}.
\]

 We define the Bayes decision rule  $\delta\left(g\right)$ for prior $\pi$ as
$
R_{\pi}\left(\delta\left(g\right),g\right)=\min_{a\in \mathcal{D}}R_{\pi}\left(a,g\right).
$
As the next step of the proof we show that for any  finitely-supported
prior $\pi$ on $\widetilde{\Gamma}$, the Bayes decision rule $\delta(g)$ is almost surely uniformly continuous in $g$ and satisfies the inequality:
\begin{align}\label{eq: continuity}
\left\Vert \delta\left(g'\right)-\delta\left(g\right)\right\Vert \leq \kappa^{-1}\left( 2\bar{L}W\cdot\left\Vert g-g'\right\Vert _{\infty}\right),
\end{align}
where $\kappa(\cdot)$ is the function from the definition of uniform convexity.

From the definition of the Bayes decision rule and
equation (\ref{eq: risk is lipshitz}) we have
$$
R_{\pi}\left(\delta\left(g'\right),g\right)-R_{\pi}\left(\delta\left(g\right),g'\right)\leq R_{\pi}\left(\delta\left(g'\right),g\right)-R_{\pi}\left(\delta\left(g'\right),g'\right)\leq \bar{L} W\left\Vert g-g'\right\Vert _{\infty}.
$$
Equation (\ref{eq: risk is lipshitz}) also implies
$
R_{\pi}\left(\delta\left(g\right),g'\right) - R_{\pi}\left(\delta\left(g\right),g\right) \leq \bar{L} W\left\Vert g-g'\right\Vert _{\infty}.
$
These inequalities result in the following statement: 
\begin{align}\label{eq: risk is lipshits in first argument}
R_{\pi}\left(\delta\left(g'\right),g\right)-R_{\pi}\left(\delta\left(g\right),g\right)\le 2\bar{L} W\cdot\left\Vert g-g'\right\Vert _{\infty}.
\end{align}
The uniform convexity of the loss function $L(a,\theta)$ results in the uniform convexity of the posterior risk function $R_{\pi}\left(a,g\right)$ in its first argument with the same $\kappa(\cdot)$.
So 
\[
R_{\pi}\left(t a+\left(1-t\right) a',g\right)\le t R_{\pi}\left(a,g\right)+\left(1-t\right) R_{\pi}\left(a',g\right)-t\left(1-t\right)\kappa\left(\left\Vert a-a'\right\Vert \right).
\]
Thus,
\[
\begin{array}{c}
R_{\pi}\left(t\delta\left(g'\right)+\left(1-t\right)\delta\left(g\right),g\right)-R_{\pi}\left(\delta\left(g\right),g\right)+t\left(1-t\right)\kappa\left(\left\Vert \delta\left(g'\right)-\delta\left(g\right)\right\Vert \right)\\
\le t\left(R_{\pi}\left(\delta\left(g'\right),g\right)-R_{\pi}\left(\delta\left(g\right),g\right)\right).
\end{array}
\]
The fact that $\delta\left(g\right)$ is a Bayes decision rule implies
that
\[
R_{\pi}\left(t\delta\left(g'\right)+\left(1-t\right)\delta\left(g\right),g\right)-R_{\pi}\left(\delta\left(g\right),g\right)\ge0.
\]
Combining this with inequality (\ref{eq: risk is lipshits in first argument}) we obtain:
\[
\kappa\left(\left\Vert \delta\left(g'\right)-\delta\left(g\right)\right\Vert \right)\le R_{\pi}\left(\delta\left(g'\right),g\right)-R_{\pi}\left(\delta\left(g\right),g\right)\leq 2\bar{L} W\cdot\left\Vert g-g'\right\Vert _{\infty}.
\]
Since function $\kappa$ is strictly increasing and thus invertible,
we arrive at (\ref{eq: continuity}). The continuity of $\kappa(\cdot)$ at zero comes from continuity of the loss function $L(a,\theta)$.

As the final step in the proof of Theorem \ref{thm: ae continuity} we show that
(\ref{eq: continuity}) holds for any admissible rule on $\widetilde{\Gamma}$. Using an extension of a result from Brown (1986),  Theorem \ref{Thm: Brown admissibility} establishes that there exists a sequence of finitely supported priors $\pi_s$ on $\widetilde{\Gamma}$ and corresponding Bayes decision rules $\delta_{\pi_s}$
such that $\delta_{\pi_s}(g)\to \delta(g)$ as $s\to \infty$ for almost every $g$.  Since $\delta_{\pi_s}(g)$ is uniformly continuous with modulus of continuity $\tilde{\kappa}$ for all $s,$ however, it follows that $\delta_{\pi_s}$ converges pointwise to a limit $\delta^*$ which is also uniformly continuous with modulus of continuity $\tilde\kappa$.  Hence, $\delta(g)=\delta^*(g)$ for almost every $g.$ $\Box$

\paragraph{Proof of Theorem \ref{thm: ae continuity1}}
First consider $\widetilde{\Gamma}\subseteq\Gamma_W$.
Following exactly the same arguments as the ones used in the proof of statement (\ref{eq: risk derivative}), we  can show that for any finitely-supported
prior $\pi$ on $\widetilde{\Gamma}$, any moment realizations $g,$ $g',$ and $g_t=g+t\cdot\tilde{g}$ for $\tilde{g}=g'-g$
\[
\frac{\partial}{\partial t}\mathbb{E}_{\pi}\left[r(\theta)|g_t\right]=\operatorname{Cov}_{\pi}\left(r(\theta),\eta_{m}(\tilde g) |g_t\right),
\]
where by the Cauchy-Schwarz inequality
\[
\|\operatorname{Cov}_{\pi}\left(r(\theta),\eta_{m}(\tilde g) |g_t\right)\|\le \bar{r}\sqrt{p}W\|g-g'\|_\infty,
\]
from which it is immediate that
$\mathbb{E}_{\pi}\left[r\left(\theta\right)|g\right]$
is Lipschitz in $g$:
\[
\left\| \mathbb{E}_{\pi}\left[r\left(\theta\right)|g\right]-\mathbb{E}_{\pi}\left[r\left(\theta\right)|g'\right]\right\| \le K\left\| g-g'\right\|_\infty.
\]
Given this Lipschitz property for Bayes posterior means, that $\delta$ is almost-everywhere Lipschitz follows by the same argument as in the proof of Theorem \ref{thm: ae continuity}. $\Box$

\paragraph{Proof of Corollary \ref{cor: Continuous}}

Immediate from Theorem \ref{thm: ae continuity}. $\Box$

\paragraph{Proof of Lemma \ref{lem: non-Lipschitz}}
Suppose that $\delta$ is both almost-surely uniformly continuous and  scale-invariant.  Consider two independent draws $g$ and $g',$ and note that by scale-invariance we have 
\[
\delta(g,\Sigma)-\delta(g',\Sigma)=\delta(c\cdot g,\Sigma)-\delta(c\cdot g',\Sigma)
\]
for all $c>0$.  However, the uniform continuity property implies  for almost every $(g,g')$ and any fixed $c,$
\[
\|\delta( g,\Sigma)-\delta( g',\Sigma)\|=\|\delta(c\cdot g,\Sigma)-\delta(c\cdot g',\Sigma)\|\le \tilde\kappa(c\cdot \|g-g'\|_\infty)
\]
with probability one.  Sending $c$ to zero leads us to the required statement. $\Box$

\paragraph{Proof of Proposition \ref{prop: bagged estimator}}

If the covariance function $\Sigma$ has a finite number of nonzero
eigenvalues, it follows that for $G\sim\mathcal{GP}(0,\Sigma)$
the process $G\left(\cdot\right)$ is a transformation of a finite-dimensional
normal random vector, so we can write $G\left(\theta\right)=A\left(\theta\right)Y$
for $Y\in\mathbb{R}^{q}$ a standard normal random
vector and $A\left(\cdot\right)$ a matrix-valued function that depends
on $\Sigma$. Correspondingly, the RKHS $\mathcal{H}$ can be written
as $\left\{ A\left(\cdot\right)x:x\in\mathbb{R}^{q}\right\} .$

Combining these observations, we can write $g\left(\cdot\right)=A\left(\cdot\right)y$
for $y\sim N\left(x,I\right),$ and any estimator $\delta\left(g\right)$
can be equivalently expressed as $\gamma\left(y\right)=\delta\left(A\left(\cdot\right)y\right).$
For $\upsilon$ a standard normal random vector and $\zeta$ as defined
in the main text, we likewise have the equality
\[
\delta^{B}(g)\equiv\mathbb{E}\left[\delta\left(g+\zeta\right)|g\right]=\mathbb{E}\left[\gamma\left(y+\upsilon\right)|y\right]\equiv\gamma^{B}\left(y\right)
\]
for the bagged estimators.
Since $\Sigma\left(\theta,\tilde{\theta}\right)=A\left(\theta\right)A\left(\tilde{\theta}\right)'$
while $\Sigma$ is continuous and $\Theta$ is compact, the largest
singular value of $A\left(\theta\right)$, $\sigma_{\max}\left(A\left(\theta\right)\right)$,
is uniformly bounded. For $\bar{\sigma}=\sup_{\theta\in\Theta}\sigma_{\max}\left(A\left(\theta\right)\right)$,
\[
\sup_{\theta\in\Theta}\left\Vert g\left(\theta\right)-\tilde{g}\left(\theta\right)\right\Vert \le\sup_{\theta\in\Theta}\sigma_{\max}\left(A\left(\theta\right)\right)\left\Vert y-\tilde{y}\right\Vert 
\le\bar{\sigma}\left\Vert y-\tilde{y}\right\Vert.
\]
Hence,
it suffices to show that $\gamma^{B}\left(y\right)$ is Lipschitz
in $y$.

Note that for
$\varphi$ the standard (multivariate) normal density,
\[
\gamma^{B}\left(y\right)=\int\gamma\left(y+\upsilon\right)\varphi\left(\upsilon\right)d\upsilon=\int\gamma\left(\upsilon\right)\varphi\left(\upsilon-y\right)d\upsilon.
\]
Hence, if we let $y_{t}=y+t\cdot\tilde{y},$ we have
\[
\frac{\partial}{\partial t}\left.\gamma^{B}\left(y_{t}\right)\right|_{t=0}=\int\gamma\left(\upsilon\right)\frac{\partial}{\partial t}\left.\varphi\left(\upsilon-y_{t}\right)\right|_{t=0}d\upsilon
\]
\[
=\int\gamma\left(\upsilon\right)\left(\upsilon-y\right)'\tilde{y}\varphi\left(\upsilon-y\right)d\upsilon=\operatorname{Cov}_{\upsilon \sim N\left(y,I\right)}\left(\gamma\left(\upsilon\right),\upsilon'\tilde{y}\right),
\]
where $\operatorname{Cov}_{\upsilon \sim N\left(y,I\right)}\left(\gamma\left(\upsilon\right),\upsilon'\right)$
denotes the covariance of $\gamma\left(\upsilon\right)$ and $\upsilon$
when $\upsilon\sim N\left(y,I\right)$. Note, however, that since the
range of $\gamma\left(\upsilon\right)$ is contained in $\mathcal{A}$, Cauchy-Schwarz
implies that for $\bar{a}=\sup_{a\in\mathcal{A}}\|a\|,$ 
\[
\left\|\frac{\partial}{\partial t}\left.\gamma^{B}\left(y_{t}\right)\right|_{t=0}\right\|=\left\Vert \operatorname{Cov}_{\upsilon \sim N\left(y,I\right)}\left(\gamma\left(\upsilon\right),\upsilon'\tilde{y}\right)\right\Vert \le\bar{a}\sqrt{p}\left\Vert \tilde{y}\right\Vert ,
\]
which completes the proof. $\Box$

\paragraph{Proof of Lemma \ref{lem: QB lipchitz in Q}}
Let $Q(\theta)=Q(\theta|g)$ and $Q'(\theta)=Q(\theta|g')$, and consider $$Q_t(\theta)=Q(\theta)+t(Q'(\theta)-Q(\theta))=Q(\theta)+t\cdot\Delta(\theta)$$ 
for $\Delta(\theta)=Q'(\theta)-Q(\theta)$. Let us write $\mathbb{E}^{QB}_\pi[\cdot|Q]$ for the expectation under the quasi-Bayes posterior distribution, which draws $\theta$ from the distribution with density $\frac{\exp(-\frac{1}{2}Q(\theta))}{\int \exp(-\frac{1}{2}Q(\theta))d\pi(\theta)}$ relative to $\pi$, and define 
$\operatorname{Cov}^{QB}_\pi(\cdot,\cdot|Q)$ analogously. Note that $\delta^{QB}_{\pi}(g)=\mathbb{E}^{QB}_{\pi}[r(\theta)|Q(\cdot|g)]$, and that
\begin{align*}
\frac{\partial}{\partial t}\mathbb{E}^{QB}_{\pi}[r(\theta)|Q_t]=\frac{\partial}{\partial t}\left[\frac{\int r\left(\theta\right)\exp\left(-\frac{1}{2}Q_t(\theta)\right)d\pi\left(\theta\right)}{\int\exp\left(-\frac{1}{2}Q_t(\theta)\right)d\pi\left(\theta\right)}\right]
\\
=-\frac{1}{2}\left(\mathbb{E}^{QB}_{\pi}[r(\theta)\Delta(\theta)|Q_t]-\mathbb{E}^{QB}_{\pi}[r(\theta)|Q_t]\mathbb{E}^{QB}_{\pi}[\Delta(\theta)|Q_t]\right)\\
=-\frac{1}{2}\operatorname{Cov}^{QB}_{\pi}\left(r\left(\theta\right),\Delta\left(\theta\right)|Q_t\right)
\end{align*}
By the Cauchy-Schwarz inequality, however,
\[
\|\operatorname{Cov}^{QB}_{\pi}\left(r\left(\theta\right),\Delta\left(\theta\right)|Q_t\right)\|\leq\bar{r}\sqrt{p}\sup_\theta|\Delta(\theta)|,
\] 
which completes the proof.  $\Box$

\paragraph{Proof of Proposition \ref{prop: quasi-bayes lipschitz}}
Similar to the proof of Lemma \ref{lem: QB lipchitz in Q}, let  $\mathbb{E}^{QB}_\pi[\cdot|g]$ be the expectation under the quasi-Bayes posterior distribution, which draws $\theta$ from the distribution with density $\frac{\exp(-\frac{1}{2}Q(\theta|g))}{\int \exp(-\frac{1}{2}Q(\theta|g))d\pi(\theta)}$ relative to $\pi$, and define 
$\operatorname{Cov}^{QB}_\pi(\cdot,\cdot|g),$ $\operatorname{Var}^{QB}_\pi(\cdot|g)$ analogously.
For $g_{t}=g+t\cdot \tilde{g}$, note that by Cauchy-Schwarz
\[
\left\|\frac{\partial}{\partial t}\delta^{QB}_{\pi}(g_t)\right\|
=\left\|\frac{\partial}{\partial t}\mathbb{E}^{QB}_{\pi}\left[r\left(\theta\right)|g_{t}\right]\right\|=\frac{1}{2}\left\|\operatorname{Cov}^{QB}_{\pi}\left(r\left(\theta\right),\tilde{g}\left(\theta\right)'\Sigma\left(\theta\right)^{-1}g_t\left(\theta\right)|g_t\right)\right\|\le
\]
\[
 \frac{1}{2}\bar{r}\sqrt{p}
\sqrt{\mathbb{E}^{QB}_{\pi}\left[\left(\tilde{g}\left(\theta\right)'\Sigma\left(\theta\right)^{-1}g_t\left(\theta\right)\right)^{2}|g_t\right]}.
\]
By another application of Cauchy-Schwarz, 
\[
\left(\tilde{g}\left(\theta\right)'\Sigma\left(\theta\right)^{-1}g_t\left(\theta\right)\right)^{2}\le\tilde{g}\left(\theta\right)'\Sigma\left(\theta\right)^{-1}\tilde{g}\left(\theta\right)\cdot Q(\theta|g_t),
\]
so 
\[
\mathbb{E}^{QB}_{\pi}\left[\left(\tilde{g}\left(\theta\right)'\Sigma\left(\theta\right)^{-1}g_t\left(\theta\right)\right)^{2}|g_t\right]\le\left\Vert \tilde{g}\right\Vert _{\Sigma,\infty}^2\mathbb{E}^{QB}_{\pi}\left[Q(\theta|g_t)|g_t\right]
\]
for 
\[
\left\Vert \tilde{g}\right\Vert _{\Sigma,\infty}=\sup_{\theta\in\Theta}\sqrt{\tilde{g}\left(\theta\right)'\Sigma\left(\theta\right)^{-1}\tilde{g}\left(\theta\right)}=\sup_{\theta\in\Theta}\sqrt{Q(\theta|\tilde{g})}.
\]
Altogether, we obtain that 
\[
\left\|\frac{\partial}{\partial t}\left.\mathbb{E}^{QB}_{\pi}\left[r\left(\theta\right)|g_{t}\right]\right|_{t=0}\right\|\le\frac{1}{2}\bar{r}\sqrt{p}\left\Vert \tilde{g}\right\Vert _{\Sigma,\infty}\sqrt{\mathbb{E}^{QB}_{\pi}\left[Q(\theta|g)|g\right]}.
\]
Note, next, that 
\[
\mathbb{E}^{QB}_{\pi}\left[Q(\theta|g)|g\right]=\frac{\int Q(\theta|g)\exp\left(-\frac{1}{2}Q(\theta|g)\right)d\pi\left(\theta\right)}{\int\exp\left(-\frac{1}{2}Q(\theta|g)\right)d\pi\left(\theta\right)}.
\]
Since the function $h\left(x\right)=x\exp\left(-\frac{1}{2}x\right)$
is maximized at $x=2,$ if 
\[
\int\exp\left(-\frac{1}{2}Q(\theta|g)\right)d\pi\left(\theta\right)\ge\varepsilon,
\]
then
\[
E_{\pi}\left[Q(\theta|g)|g\right]\le 2\exp\left(-1\right)\varepsilon^{-1}.
\]
Note, however, that for $|\Theta|$ finite and $\underline{\pi}=\min_{\theta\in\Theta}\pi(\theta),$
\[
\int\exp\left(-\frac{1}{2}Q(\theta|g)\right)d\pi\left(\theta\right) =
\sum_{\theta\in\Theta} \exp\left(-\frac{1}{2}Q(\theta|g)\right)\pi\left(\theta\right)
\ge \exp\left(-\frac{1}{2}\min_{\theta\in\Theta}Q(\theta|g)\right)\underline{\pi}.
\]
Hence, for $g\in\mathcal{G}_C$ as defined in the proposition, 
\[
\left\|\frac{\partial}{\partial t}\left.\mathbb{E}^{QB}_{\pi}\left[r\left(\theta\right)|g_{t}\right]\right|_{t=0}\right\|
\le \bar{r}\sqrt{p}\left\Vert \tilde{g}\right\Vert_{\Sigma,\infty} \exp\left(\frac{1}{2}C-1\right)\underline{\pi}^{-1},
\]
which completes the proof. $\Box$

\section{Example: Discontinuous Limit-of-Bayes}\label{sec: discontinuity example}

This appendix provides an example to demonstrate that without bounds on identification strength, the pointwise limit of Bayes posterior means can be discontinuous.  

Let us continue the finite $\Theta$ special case discussed in main text, and further suppose that the parameter space consists of just two points, $\Theta=\{\theta_1,\theta_2\},$ that we have a one-dimensional moment condition ($k=1$), and that $\Sigma=I_2$. The function $m$ is thus described by two numbers -- the values at $\theta_1$ and $\theta_2$.  Consider prior $\pi_C$, supported on just two values of $(\theta,m)$, which  assigns probability $\frac{1}{2}$ to each of $\theta_1$ and $\theta_2$ and, conditional on $\theta=\theta_j$, implies that $(m(\theta_j),m(\theta_{-j}))=(0,C)$ with probability one, where $\theta_{-j}$ denotes the element of $\Theta$ other than $\theta_j$.  Suppose we are interested in estimating $r(\theta)=\theta$ under squared error loss, and note that the Bayes estimator corresponds to the posterior mean
\[
\delta_{\pi_C}(g,\Sigma)=\mathbb{E}_{\pi_C}[\theta|g]=\frac{\sum_{j=1}^2 \theta_j\exp(-\frac{1}{2}g(\theta_j)^2-\frac{1}{2}(g(\theta_{-j})-C)^2)}{\sum_{j'=1}^2\exp(-\frac{1}{2}g(\theta_{j'})^2-\frac{1}{2}(g(\theta_{-j'})-C)^2)}.
\]
For a given $g$ with $g(\theta_1)\neq g(\theta_2)$ $\mathbb{E}_{\pi_C}[\theta|g]\to {\arg\min}_{\theta\in\{\theta_1,\theta_2\}} g(\theta)$ as $C\to\infty$. The limiting estimator $\delta(g)={\arg\min}_{\theta\in\{\theta_1,\theta_2\}} g(\theta)$ is a step function, and thus discontinuous.

\section{Strong-Identification Limit Experiment}\label{sec: strong ID}

This appendix outlines the limit experiment for strongly identified GMM and discusses why our inadmissibility argument for GMM estimators under weak identification does not translate to this setting.

In strongly identified GMM models, $\theta^*$ is consistently estimable at a $\sqrt{n}$ rate.  To obtain a non-trivial decision problem in the limit we must therefore focus attention on a shrinking neighborhood of the true parameter value, reparameterizing the model in terms of $\tau^*=\sqrt{n}(\theta^*_n-\theta_0),$ for $\theta^*_n$ the true parameter value in the sample of size $n$ and $\theta_0$ a fixed parameter value around which we localize the problem.  
Under mild conditions on the moments, the limit experiment for estimation of $\tau^*$ in this setting corresponds to observing
\begin{equation}\label{eq: strong limit experiment}
g(0)\sim N(-H\tau^*,\Omega)
\end{equation}
for $H$ and $\Omega$ known, non-stochastic matrices where $H$ has full column rank, and $\tau^*\in\mathbb{R}^p$. See for instance Armstrong and Kolesar (2021).

Observing (\ref{eq: strong limit experiment}) is equivalent to observing the linear process $g(\cdot)$ where
\[
g(\cdot)=\mathcal{GP}(m,\Sigma).
\]
with $m(\tau)=H(\tau-\tau^*)$ and $\Sigma(\tau,\tilde\tau)=\Omega$, where both $H$ and $\Sigma$ are known.
Hence, the strong-identification limit experiment (\ref{eq: strong limit experiment}) can be cast as a special case of our weak-identification limit experiment (\ref{eq: GP model}) with the additional restrictions that (i) $m$ is linear with a known Jacobian and (ii) $\Sigma$ is constant.  
%Constancy of $\Sigma$ implies that the RKHS $\mathcal{H}$ in this setting is simply the set of constant functions. 

Our inadmissibility argument for GMM under weak identification relies on the fact that the GMM estimator is the same for moment realizations $g(\cdot)$ and $c\cdot g(\cdot)$ for all positive constants $c$.  In the strongly identified case, however, $c\cdot g(\cdot)$ for $c\neq 1$ has a Jacobian different from $H$, and so lies outside the support of the process $g(\cdot).$  If we instead rescale the moments only at a given point $\tau,$ note that the GMM estimator for $\tau^*$ with weighting matrix  $\Xi$ can be written as $\hat\tau=\tau-(H'\Xi H)^{-1}H'\Xi g(\tau).$  Hence, in the strongly identified case GMM is not invariant to changing the scale of $g(\tau)$.

Indeed, a stronger statement is true: if we consider a flat prior on $\tau^*,$ the posterior mean is equal to the efficient GMM estimator with weighting matrix $\Xi=\Omega^{-1}.$  Hence, in the strongly-identified limit experiment the efficient GMM estimator can be written as the limit of a sequence of Bayes posterior means for proper priors, and our inadmissibility argument does not apply.

\section{Additional Simulation Results}

This section reports additional simulation results to complement those reported in the main text.  Tables \ref{tab: Simulation Results, k} and \ref{tab: Simulation Results, k, corr} report average normalized mean squared error, for the IV and correlation coefficients respectively, when we bin specifications based on the number of instruments rather than the effective F statistic.  We see that the gains for the alternative estimators are largest in the just-identified case, while the performance of the GMM and alternative estimators tends to be closer in the over-identified specifications.

Tables \ref{tab: Simulation Results,  varying space} and \ref{tab: Simulation Results,  corr, varying space} report normalized root mean squared error (based on 1,000 simulation draws) averaged across all specifications when we set the parameter space $\Theta_s$ in specification $s$ equal to $\left[\pm c\cdot\frac{\sigma_{uv}}{\sigma_v^2}\right]$ for $c\in\{20,40,60\}.$  As these results show, the qualitative messages from our simulations are unchanged across these alternative parameter spaces.  Specifically, Table \ref{tab: Simulation Results,  varying space} reports results for the IV coefficient, and we see that while the root mean squared error for all estimators grows as we widen the bounds on the parameter space, in every case bagged two-stage least squares performs the best, followed by bagged continuously updating GMM, then quasi-Bayes with an invariant prior, then the two GMM estimators, and finally quasi-Bayes with a flat prior.  Table \ref{tab: Simulation Results,  corr, varying space} reports results for the correlation coefficient and shows, as expected, that the bounds of the parameter space make almost no difference in this case.

Tables \ref{tab: Simulation Results, TSLS, F} and \ref{tab: Simulation Results, TSLS, k} drop boundedness of the parameter space altogether and report the mean squared error of two-stage least squares and bagged two-stage least squares estimators for the IV coefficient when we limit attention to specifications with $k\ge3$.  Table \ref{tab: Simulation Results, TSLS, F} bins the specifications based on the effective F-statistic, while Table \ref{tab: Simulation Results, TSLS, k} bins them based on the number of instruments. We see that in these specifications bagging makes very little difference: while the bagged estimator performs slightly better in some cases, and slightly worse in others, the overall performance difference between the two estimators is minimal.  This is consistent with the results in Table \ref{tab: Simulation Results, k}, which show that the performance gains for bagged two stage least squares are coming from the specifications with $k=1$ and $k=2.$

\clearpage

\begin{table}
\begin{tabular}{ccccc}
 & $k=1$  & $k=2$  & $k=3$  & $k\ge4$\tabularnewline
\hline 
\hline 
$\begin{array}{c}
\text{Two-Stage}\\
\text{ Least Squares}
\end{array}$  & 1.84 & 1.06 & 1.01 & 0.98\tabularnewline
\hline 
$\begin{array}{c}
\text{Continuously}\\
\text{ Updating GMM}
\end{array}$  & 1.84 & 1.12 & 1.04 & 1.21\tabularnewline
\hline 
$\begin{array}{c}
\text{Bagged Two-Stage}\\
\text{ Least Squares}
\end{array}$  & 1.12 & 1.04 & 0.99 & 1.01\tabularnewline
\hline 
$\begin{array}{c}
\text{Bagged Continuously}\\
\text{ Updating GMM}
\end{array}$  & 1.12 & 1.08 & 1.03 & 1.00\tabularnewline
\hline 
$\begin{array}{c}
\text{Quasi-Bayes,}\\
\text{ Flat Prior}
\end{array}$  & 1.94 & 1.20 & 1.13 & 1.27\tabularnewline
\hline 
$\begin{array}{c}
\text{Quasi-Bayes,}\\
\text{Invariant Prior}
\end{array}$  & 1.06 & 1.07 & 1.04 & 1.22\tabularnewline
\hline 
$\text{Number of Specifications}$  & 31 & 20 & 30 & 40\tabularnewline
\end{tabular}

\caption{Performance of IV coefficient estimators in Andrews et al. (2019) specifications. Entries
correspond to the root mean squared error normalized by the standard
error in the Andrews et al. (2019) data, $\sqrt{\mathbb{E}_{s}[(\delta(g)-\theta_{s}^{*})^{2}]}/\sigma_{s}^{*}$,
averaged across specifications. Columns correspond to varying degrees of over-identification.\label{tab: Simulation Results, k}}
\end{table}

\begin{table}
\begin{tabular}{ccccc}
 & $k=1$  & $k=2$  & $k=3$  & $k\ge4$\tabularnewline
\hline 
\hline 
$\begin{array}{c}
\text{Two-Stage}\\
\text{ Least Squares}
\end{array}$  & 1.20 & 1.06 & 1.00 & 1.01\tabularnewline
\hline 
$\begin{array}{c}
\text{Continuously}\\
\text{ Updating GMM}
\end{array}$  & 1.20 & 1.10 & 1.01 & 1.18\tabularnewline
\hline 
$\begin{array}{c}
\text{Bagged Two-Stage}\\
\text{ Least Squares}
\end{array}$  & 1.00 & 1.07 & 0.98 & 1.06\tabularnewline
\hline 
$\begin{array}{c}
\text{Bagged Continuously}\\
\text{ Updating GMM}
\end{array}$  & 1.00 & 1.10 & 1.00 & 1.01\tabularnewline
\hline 
$\begin{array}{c}
\text{Quasi-Bayes,}\\
\text{ Flat Prior}
\end{array}$  & 1.38 & 1.19 & 1.04 & 1.19\tabularnewline
\hline 
$\begin{array}{c}
\text{Quasi-Bayes,}\\
\text{Invariant Prior}
\end{array}$  & 0.99 & 1.11 & 1.00 & 1.17\tabularnewline
\hline 
$\text{Number of Specifications}$  & 31 & 20 & 30 & 40\tabularnewline
\end{tabular}
\caption{Performance of correlation coefficient estimators in Andrews et al. (2019) specifications. Entries
correspond to the root mean squared error normalized by the standard
error in the Andrews et al. (2019) data, $\sqrt{\mathbb{E}_{s}[(\delta(g)-r(\theta_{s}^{*}))^{2}]}/\sigma_{r,s}^{*}$,
averaged across specifications. Columns correspond to varying degrees of over-identification.\label{tab: Simulation Results, k, corr}}
\end{table}

\begin{table}
\begin{tabular}{cccc}
 & $\Theta_{s}=\left[\pm20\frac{\sigma_{uv,s}}{\sigma_{v,s}^{2}}\right]$  & $\Theta_{s}=\left[\pm40\frac{\sigma_{uv,s}}{\sigma_{v,s}^{2}}\right]$  & $\Theta_{s}=\left[\pm60\frac{\sigma_{uv,s}}{\sigma_{v,s}^{2}}\right]$ \tabularnewline
\hline 
\hline 
$\begin{array}{c}
\text{Two-Stage}\\
\text{ Least Squares}
\end{array}$  & 1.22 & 1.34 & 1.42\tabularnewline
\hline 
$\begin{array}{c}
\text{Continuously}\\
\text{ Updating GMM}
\end{array}$  & 1.30 & 1.42 & 1.51\tabularnewline
\hline 
$\begin{array}{c}
\text{Bagged Two-Stage}\\
\text{ Least Squares}
\end{array}$  & 1.04 & 1.06 & 1.07\tabularnewline
\hline 
$\begin{array}{c}
\text{Bagged Continuously}\\
\text{ Updating GMM}
\end{array}$  & 1.05 & 1.07 & 1.09\tabularnewline
\hline 
$\begin{array}{c}
\text{Quasi-Bayes,}\\
\text{ Flat Prior}
\end{array}$  & 1.39 & 1.58 & 1.70\tabularnewline
\hline 
$\begin{array}{c}
\text{Quasi-Bayes,}\\
\text{Invariant Prior}
\end{array}$  & 1.11 & 1.14 & 1.15\tabularnewline
\hline 
$\text{Number of Specifications}$  & 121 & 122 & 122\tabularnewline
\end{tabular}

\caption{Performance of IV coefficient estimators in Andrews et al. (2019)
specifications based on 1000 simulation draws. Entries correspond
the root mean squared error normalized by the standard error in the
Andrews et al. (2019) data, $\sqrt{\mathbb{E}_{s}[(\delta(g)-\theta_{s}^{*})^{2}]}/\sigma_{s}^{*}$,
averaged across all specifications. Columns vary the size of the parameter
space.\label{tab: Simulation Results,  varying space}}
\end{table}

\begin{table}
\begin{tabular}{cccc}
 & $\Theta_{s}=\left[\pm20\frac{\sigma_{uv,s}}{\sigma_{v,s}^{2}}\right]$ & $\Theta_{s}=\left[\pm40\frac{\sigma_{uv,s}}{\sigma_{v,s}^{2}}\right]$ & $\Theta_{s}=\left[\pm60\frac{\sigma_{uv,s}}{\sigma_{v,s}^{2}}\right]$\tabularnewline
\hline 
\hline 
$\begin{array}{c}
\text{Two-Stage}\\
\text{ Least Squares}
\end{array}$ & 1.07 & 1.07 & 1.07\tabularnewline
\hline 
$\begin{array}{c}
\text{Continuously}\\
\text{ Updating GMM}
\end{array}$ & 1.12 & 1.12 & 1.12\tabularnewline
\hline 
$\begin{array}{c}
\text{Bagged Two-Stage}\\
\text{ Least Squares}
\end{array}$ & 1.03 & 1.03 & 1.03\tabularnewline
\hline 
$\begin{array}{c}
\text{Bagged Continuously}\\
\text{ Updating GMM}
\end{array}$ & 1.02 & 1.03 & 1.03\tabularnewline
\hline 
$\begin{array}{c}
\text{Quasi-Bayes,}\\
\text{ Flat Prior}
\end{array}$ & 1.20 & 1.21 & 1.20\tabularnewline
\hline 
$\begin{array}{c}
\text{Quasi-Bayes,}\\
\text{Invariant Prior}
\end{array}$ & 1.07 & 1.08 & 1.08\tabularnewline
\hline 
$\text{Number of Specifications}$ & 121 & 122 & 122\tabularnewline
\end{tabular}

\caption{Performance of correlation coefficient estimators in Andrews et al.
(2019) specifications based on 1000 simulation draws. Entries correspond
the root mean squared error normalized by the standard error in the
Andrews et al. (2019) data, $\sqrt{\mathbb{E}_{s}[(\delta(g)-r(\theta_{s}^{*}))^{2}]}/\sigma_{r,s}^{*}$,
averaged across all specifications. Columns vary the size of the parameter
space.\label{tab: Simulation Results,  corr, varying space}}
\end{table}

\begin{table}
\begin{tabular}{ccccc}
 & $\mathbb{E}_{s}\left[F\right]\le10$  & $10<\mathbb{E}_{s}\left[F\right]\le20$  & $20<\mathbb{E}_{s}\left[F\right]\le50$  & $50<\mathbb{E}_{s}\left[F\right]$\tabularnewline
\hline 
\hline 
$\begin{array}{c}
\text{Two-Stage}\\
\text{ Least Squares}
\end{array}$  & 0.98 & 1.01 & 1.01 & 1.00\tabularnewline
\hline 
$\begin{array}{c}
\text{Bagged Two-Stage}\\
\text{ Least Squares}
\end{array}$  & 1.00 & 0.98 & 0.99 & 0.99\tabularnewline
\hline 
$\text{Number of Specifications}$  & 42 & 4 & 15 & 9\tabularnewline
\end{tabular}

\caption{Estimator performance in Andrews et al. (2019) specifications with $k\ge3$ and an unbounded parameter space. Entries
correspond to the root mean squared error normalized by the standard
error in the Andrews et al. (2019) data, $\sqrt{\mathbb{E}_{s}[(\delta(g)-\theta_{s}^{*})^{2}]}/\sigma_{s}^{*}$,
averaged across specifications. Columns correspond to ranges of values
for the average effective first-stage F statistic of Montiel-Olea
and Pflueger (2013).\label{tab: Simulation Results, TSLS, F}}
\end{table}

\begin{table}
\begin{tabular}{ccc}
 & $k=3$  & $k\ge4$ \tabularnewline
\hline 
\hline 
$\begin{array}{c}
\text{Two-Stage}\\
\text{ Least Squares}
\end{array}$  & 1.01 & 0.98\tabularnewline
\hline 
$\begin{array}{c}
\text{Bagged Two-Stage}\\
\text{ Least Squares}
\end{array}$  & 0.99 & 1.01\tabularnewline
\hline 
$\text{Number of Specifications}$  & 30 & 40\tabularnewline
\end{tabular}

\caption{Estimator performance in Andrews et al. (2019) specifications. Entries
correspond to the root mean squared error normalized by the standard
error in the Andrews et al. (2019) data, $\sqrt{\mathbb{E}_{s}[(\delta(g)-\theta_{s}^{*})^{2}]}/\sigma_{s}^{*}$,
averaged across specifications. Columns correspond to varying degrees of over-identification.\label{tab: Simulation Results, TSLS, k}}
\end{table}

\end{appendix}
\end{document}